\documentclass[%
 reprint,
 amsmath,amssymb,
 aps,
prl,
]{revtex4-2}

\usepackage{braket}
\usepackage{mathbbol}
\usepackage{graphicx}
\usepackage{dcolumn}
\usepackage{bm}

\begin{document}

\preprint{APS/123-QED}

\title{Integrated Backward Second-Harmonic Generation Through Optically Induced Quasi-Phase Matching}

\author{Ozan Yakar$^1$}
\author{Edgars Nitiss$^1$}%
\author{Jianqi Hu$^1$}%
\author{Camille-Sophie Br\`{e}s$^{1,}$}%
 \email{camille.bres@epfl.ch}
\affiliation{$^1$\'Ecole Polytechnique Fédérale de Lausanne (EPFL), Photonic Systems Laboratory (PHOSL), Lausanne CH-1015, Switzerland }

\date{\today}

\begin{abstract}
Quasi-phase-matching for efficient backward second-harmonic generation (BSHG) requires sub-$\rm\mu$m poling periods, a non-trivial fabrication feat. For the first time, we report integrated first-order quasi-phase-matched BSHG enabled by seeded all-optical poling. The self-organized grating inscription circumvents all fabrication challenges. We compare backward and forward processes and explain how grating period influences the conversion efficiency. These results showcase unique properties of the coherent photogalvanic effect and how it can bring new nonlinear functionalities to integrated photonics.
\end{abstract}

\maketitle

The efficiency of nonlinear wave mixing processes is clamped by the mismatch of momenta due to dispersion. Quasi-phase-matching (QPM) \cite{armstrong1962interactions} is a technique used to compensate the momentum mismatch and recover the efficiency of the parametric conversion through the periodic reversal of nonlinear susceptibility in the medium. Integrated waveguiding structures are ideal candidates for nonlinear optical processes as they offer added flexibility for dispersion engineering, higher light intensities under reduced power and access to a wide range of materials. In integrated platforms, QPM has been realized by domain inversion in lithium niobate on insulator ferroelectric waveguides through electric-field poling \cite{wang2018ultrahigh}, with p-n junctions in silicon \cite{timurdogan2017electric}, or by optically inducing a periodic electric field by all-optical poling (AOP) enabled by the photogalvanic effect. AOP, first extensively studied in optical fibers \cite{osterberg1986dye, baskin1988coherent, zel1989interference, sipeol1991, margulis1995imaging, dianov1995photoinduced}, was recently demonstrated in both silicon nitride (Si$_3$N$_4$) waveguides \cite{billat2017large,porcel2017photo,hickstein2019self} and microresonators \cite{lu2021efficient,nitiss2021optically,hu2022photo}. During AOP, multi-photon absorption interference leads to the inscription of periodic DC field automatically satisfying QPM condition for the participating waves. The obtained QPM grating and effective second-order susceptibility ($\chi^{(2)}$) overcome the typical trade-off between device functionality/performance and fabrication complexity. This effective $\chi^{(2)}$ brings added functionalities to Si$_3$N$_4$, an already mature linear platform and heavily exploited for third-order ($\chi^{(3)}$) nonlinear effects \cite{ye2021overcoming, ayan2022polarization,levy2011harmonic,grassani2019mid,kippenberg2018dissipative}. As such, the demonstration of broadband second-harmonic generation (SHG) \cite{hickstein2019self, nitiss2020broadband}, difference-frequency generation \cite{sahin2021difference} and spontaneous parametric down-conversion \cite{dalidet2022near} were reported in Si$_3$N$_4$ waveguides.

\begin{figure}
    \centering
    \includegraphics[width=1\linewidth]{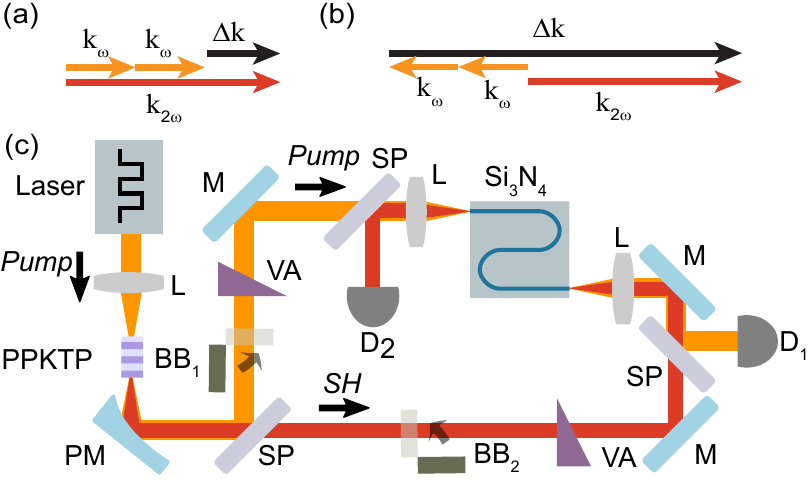}
    \caption{Illustration of momentum conservation of the a) forward and b) backward SHG processes. c) AOP setup for inducing BSHG QPM. Pulsed light is sent to PPKTP after which pump and seed SH are collimated in the parabolic mirror. Pump and SH are split and coupled through opposite ends of the waveguide, with their path lengths matched. BB$_{1,2}$: beam block for pump and SH, respectively, VA: variable attenuator, M: mirror, SP: short-pass dichroic mirror (cutoff at $1100$ nm), L: lens, PM: parabolic mirror, D$_{1,2}$: detector.}
    \label{fig:setup}
\end{figure}

Most of the QPM work, in any platform, has been devoted to forward second-harmonic generation (FSHG), where the pump and the generated second harmonic (SH) propagate in the same direction (Fig. \ref{fig:setup}a). The required QPM period, $\Lambda_{\rm FSHG}$ is determined by the momentum mismatch between the pump and its SH given by $\Lambda_{\rm FSHG}= \frac{2\pi}{\Delta k}=\frac{\lambda}{2|n_{2\omega}- n_{\omega}|}$, where $\Delta k$ is the difference in wavevectors, $\lambda$ is the pump wavelength, $n_{2\omega}$ and $n_{\omega}$ are the effective indices at the SH and pump, respectively. Typically the necessary periods, in both crystals and waveguides, are in the order of a few microns. The fabrication of external structures of such dimensions, as required for electric-field poling, is possible. However, sub-micron QPM periods necessary for mirrorless optical parametric oscillators \cite{canalias2007mirrorless}, spontaneous parametric down-conversion with very narrow spectrum \cite{luo2020counter,liu2021observation}, or backward second-harmonic generation (BSHG) \cite{dalessandro1997nonlinear}, where the pump and SH waves travel in opposite directions (Fig. \ref{fig:setup}b), face significant fabrication challenges \cite{boes2021efficient, krasnokutska2021submicron}. As such, while BSHG has been studied theoretically for several configurations \cite{conforti2008pulse, Iliew2010Huge}, there have been limited experimental demonstrations. BSHG has been achieved in bulk materials such as periodically poled lithium niobate (PPLN) \cite{busacca2014backward} and potassium titanyl phosphate (KTP) using high-order QPM \cite{Gu1999Backward, canalias2005backward, Mutter2022Third}, or stacked metasurfaces \cite{stolt2021backward} with first-order QPM. There is also a continuous effort to achieve BSHG in integrated photonics. Current demonstrations include using plasmonic structures with negative refractive index materials relying on perfect phase-matching \cite{lan2015backward, liu2018backward} as well as PPLN waveguide with higher-order QPM \cite{busacca2014backward}.

In this work, for the first time, we demonstrate AOP induced first-order QPM gratings with sub-micron periods ($\Lambda_{\rm BSHG}=\frac{\lambda}{2|n_{2\omega}+ n_{\omega}|}$ see Fig. \ref{fig:setup}b) enabling BSHG with on-chip conversion efficiency (CE), defined as $\eta_{2\omega}=P_{2\omega}/P_{\omega}^2$, of $1.2\times10^{-4}\,{\rm \%/W}$, which in the C-band is comparable with the highest CE value reached in any platform \cite{busacca2014backward}. We achieve this by leveraging the self-organization properties of the optically written gratings, and by seeding the AOP process with counter-propagating coherent pump and SH seed light, bypassing the complex fabrication steps utilized in other platforms. We confirm that BSHG allows for narrow bandwidth SHG as well as very high thermal sensitivity compared to its bandwidth. We also explain how the grating period values affect the achievable CE. 

The AOP process results in a grating with a periodicity of $\chi^{(2)}$ automatically compensating the wavevector mismatch of the interfering coherent waves responsible for the effect. During the process, photogalvanic current ($j_{ph}$), proportional to $(E_{\omega}^*)^2 E_{2\omega}e^{i\Delta k z}+c.c.$, leads to charge separation and therefore inscription of a DC electric field with periodic modulation satisfying QPM condition. Here, $E_{\omega}$ and $E_{2\omega}$ are the complex electric field amplitudes of pump and SH fields, respectively, $z$ is the propagation distance, $^*$ and $c.c.$ denote complex conjugate, $\Delta k$ is the difference of the wavevectors of pump and SH (Fig. 1a and b). The inscribed field saturates due to photoconductivity ($\sigma$). As such, FSHG can be obtained by solely launching a pump wave in a waveguide or by a seeded process \cite{krol1991seeded,yakar2022coherent} where both the pump and its SH are simultaneously coupled to the waveguide. By injecting the pump and its SH from opposite sides of the waveguide, a QPM grating for BSHG could therefore be optically inscribed. The experimental setup for poling using two counter-propagating beams is shown in Fig. \ref{fig:setup}c. We carried out the demonstration in an integrated Si$_3$N$_4$ waveguide buried in SiO$_2$ fabricated by LIGENTEC SA using its AN800 platform. The waveguide has a cross-section of $1.3\,{\rm \mu m}$$\times$$ 0.8\,{\rm \mu m}$ and is folded in 9 meanders for a total length of $4.3\,{\rm cm}$, including input and output tapers. Backward-seeded AOP is initiated by $1550\, {\rm nm}$ light together with its SH, externally generated using a nonlinear crystal - periodically poled KTP (PPKTP). The 1550 nm pump was first shaped into a $5\,{\rm ns}$ pulse train at $1\,{\rm MHz}$ repetition rate and was amplified to reach peak power up to $8.6\,{\rm W}$ in the waveguide. The pump and its SH  are collimated by a parabolic mirror, split and coupled through opposite ends of the waveguide in TE polarization. The power of both the forward pump and the backward SH seed can be controlled using variable attenuators. During this process, the optical path lengths are matched in order to ensure the overlap of pump and seed SH pulses inside the waveguide, as to enable grating inscription. It should be noted that the pulses, when not temporally overlapping inside the waveguide, can contribute to grating erasal. In our case we estimate that the inscription-to-erasure time ratio is around 35:1. The forward pump is monitored by detector D$_1$ while SH is collected by detector D$_2$ after a dichroic mirror. Initially, seeding SH is blocked with a beam block (BB$_2$) and no backward SH is observed. During seeded AOP, BB$_2$ is kept open, and is occasionally shut for a short period of time as to record the dynamic evolution of BSHG in the waveguide. Once saturation is reached, the performance of the device such as maximum BSH CE and QPM bandwidth is quantified using CW pump light. 

\begin{figure}
    \centering
    \includegraphics[width=1\linewidth]{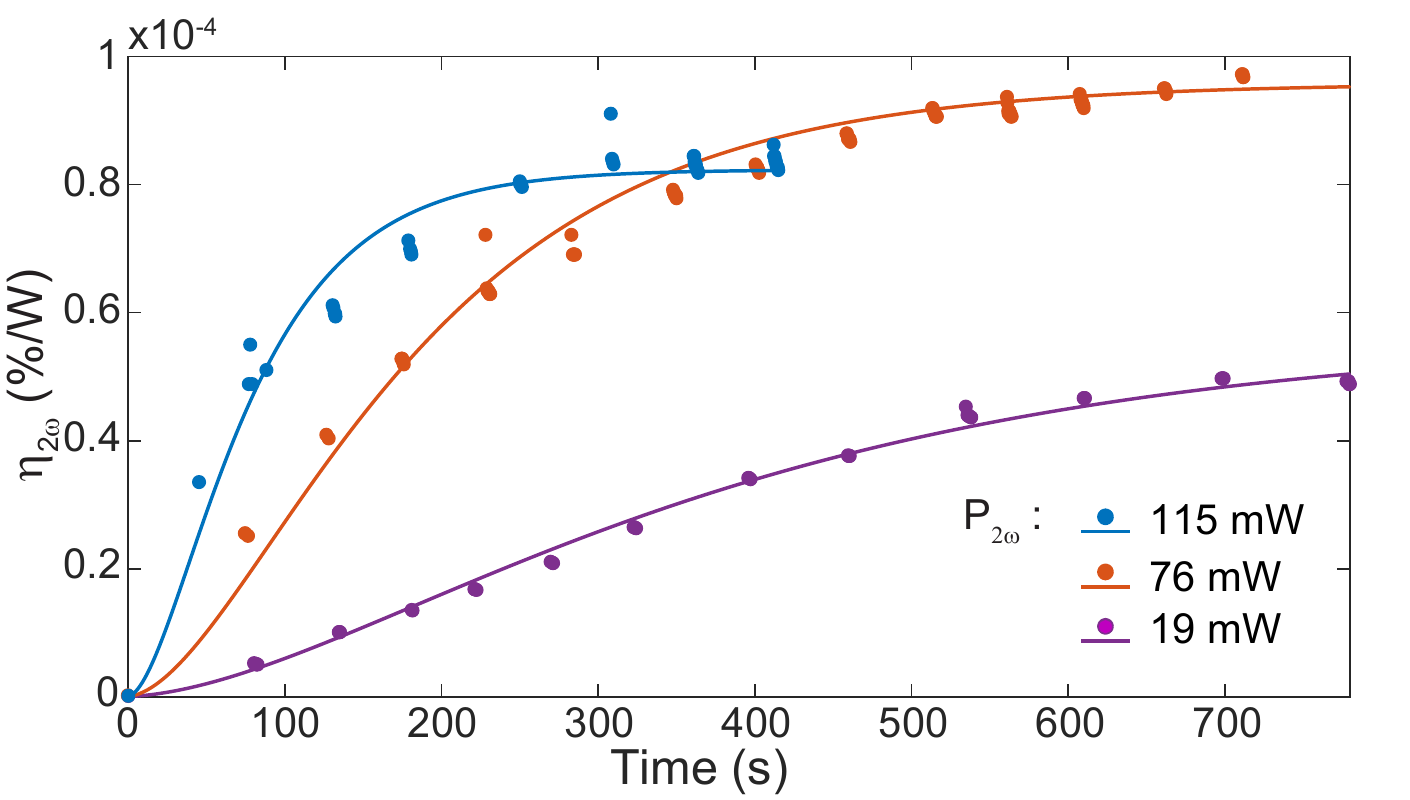}
    \caption{Experimental BSHG CE during AOP (dotted) and fit (solid line) according to Ref. \cite{yakar2022coherent} for a constant peak pump power of 8.6 W and varying peak SH seed power inside the waveguide.
    }
    \label{fig:fitgrowth}
\end{figure}

\begin{figure*}[htb]
    \centering
    \includegraphics[width=1\linewidth]{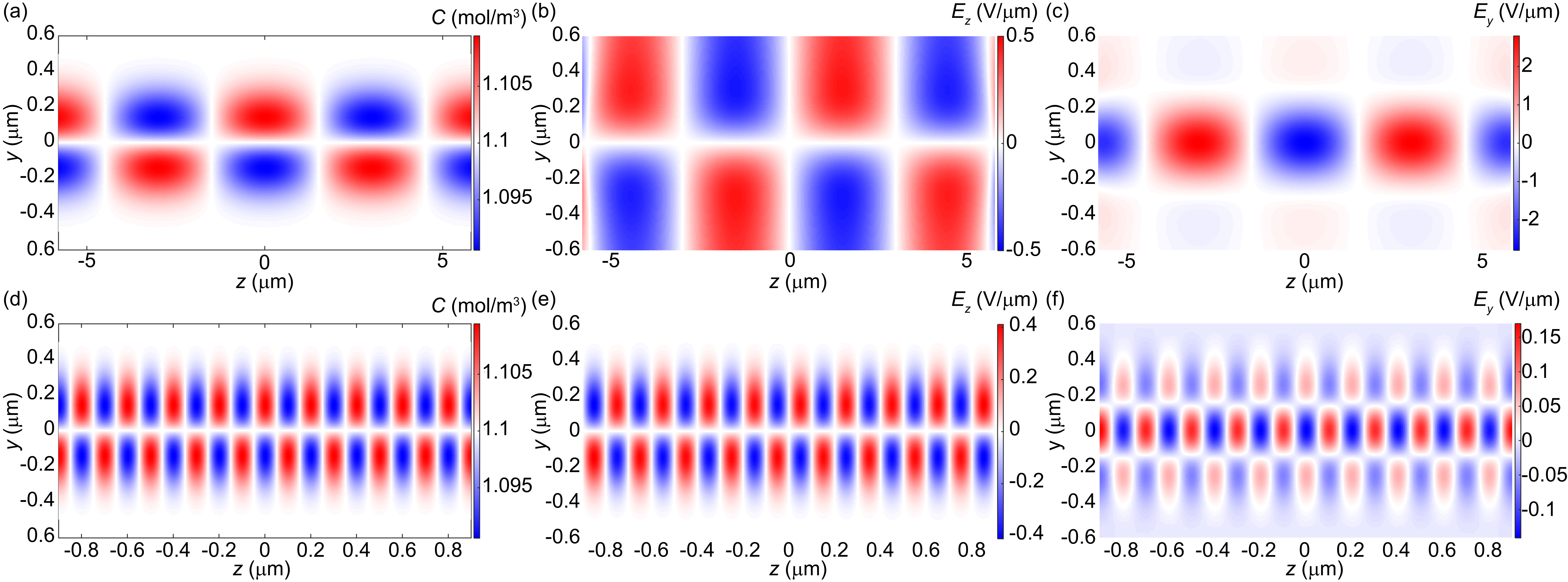}
    \caption{The effect of poling period on the inscribed DC electric field is shown. A periodically oscillating charge concentration along the waveguide $C$ (related to charge density $\rho=qN_AC$ where q is the electronic charge \cite{zabelich2022linear} and $N_A$ is the Avogadro constant) and resulting longitudinal ($E_z$) and transverse ($E_y$) components of the electric field are simulated for a period $6\,\rm \mu m$ (a-c) and $0.2\,\rm \mu m$ (d-f), respectively.}
    \label{fig:efieldsim}
\end{figure*}

The poling traces for three different SH powers and constant pump power of 8.6 W are presented in Fig. \ref{fig:fitgrowth}(a). First, we clearly measure SH being generated backward. Second, the initial growth rate, ($d{\eta_{2\omega}}/dt|_{t=0}$) increases with seed SH power. While the maximum CE also initially increases, it eventually saturates owing to the enhanced $\sigma$ \cite{yakar2022coherent}. Evidently, the resulting CEs for BSHG are close to 2 orders of magnitude lower than for FSHG when similar pump and SH seed power levels are used. The main factor controlling the efficiency of AOP is the initial growth rate, which correlates to how the current is translated to the electric field. Compared to forward AOP, $d{\eta_{2\omega}}/dt|_{t=0}$, numerically fitted following high-seed approximation \cite{yakar2022coherent}, is here found to be close to 100 times less. This is explained by the small grating period and the closeness of the separated charges, leading to screening. As a consequence, the electric field is weaker and the growth rate is reduced compared to larger grating periods even under identical photogalvanic current. Using Eq. (S21) in Supplementary Information 3, the simulated initial growth rate with respect to the grating period is shown in Fig. S1. It can be seen that the initial growth rate is reduced by approximately 10 times for backward-wave SHG (BWSHG) (when the two SH photons propagate in different directions) and 100 times for BSHG due to the screening of charges.

To confirm the behavior of the electric field as a function of the grating period, without loss of generality, let us assume that within the waveguide there is an equal number of mobile positive charge carriers and immobile negative charge carriers evenly distributed \cite{zabelich2022linear}. After AOP the positive mobile charge carrier density inside the waveguide can be described as $\rho(x,y,z)=\rho(x,y) \cos(\Delta k z) + \rho_0(x,y,z)$ where $\rho_0$ is constant. While the total number of charges in the waveguide does not change, the separated charge carriers lead to an induced local electric field. Hence, it can be observed that $\iint dx dy \rho(x,y) = 0$ as the charges are generated in pairs of opposite signs. Using Coulomb's law, the y-component of the electric field ($E_y$) at an arbitrary position (x,y,z) becomes
\begin{widetext}
\begin{equation}
    E_y(\Delta k; x,y,z) = \frac{2\Delta k\cos(\Delta k z)}{4\pi\epsilon} \iint dx^\prime dy^\prime  \rho(x^\prime+x,y^\prime+y) y\frac{K_1(\Delta k \sqrt{x^2+y^2})}{\sqrt{x^2+y^2}}~,
\end{equation}
\end{widetext}
where $K_1$ is the first-order modified Bessel function of the second kind (see Supplementary Information 1). As $K_1$ is an exponentially decaying function, $E_y$ will decay with increasing $\Delta k$, namely, with decreasing grating period. This is due to the screening of electric fields. The effect of the period on the achievable electric fields was simulated using COMSOL Multiphysics software for a charge distribution of $C=C_{\rm bulk}(1+ye^{-y^2/\gamma^2}\cos(\Delta k z))$ using the material constants and physical model described in Ref. \cite{zabelich2022electric}.
We showcase simulation results for two specific cases with periodicity $\Lambda$ of charge spatial modulation being 6 $\rm \mu m$ and 0.2 $\rm \mu m$. Gratings with such period can satisfy the QPM condition for FSHG and BSHG, respectively, as will be shown below. 
In Fig. \ref{fig:efieldsim}(a)-(c) the simulated charge distribution $C$, longitudinal electric field $E_z$ and transverse electric field $E_y$ are shown for the case when $\Lambda$=6 $\rm \mu m$. It is important to note that the transverse field strength $E_y$ determines the effective second-order nonlinearity $\chi^{(2)}_{\rm eff}$ and, hence the CE. In Fig. \ref{fig:efieldsim}(d)-(f) we also display the $C$, $E_z$ and $E_y$ with $\Lambda$=0.2 $\rm \mu m$. By comparing the $E_y$ in both cases, it can be recognized that the magnitude of $E_y$ reduces approximately ten-fold because of screening when the period is decreased from 6 $\rm \mu m$ to 0.2 $\rm \mu m$. Notably, the longitudinal field $E_z$ is almost the same for both cases as can be seen from Fig. \ref{fig:efieldsim}(b) and (e). 

\begin{figure}[h!]
  \centering
  \includegraphics[width=1\linewidth]{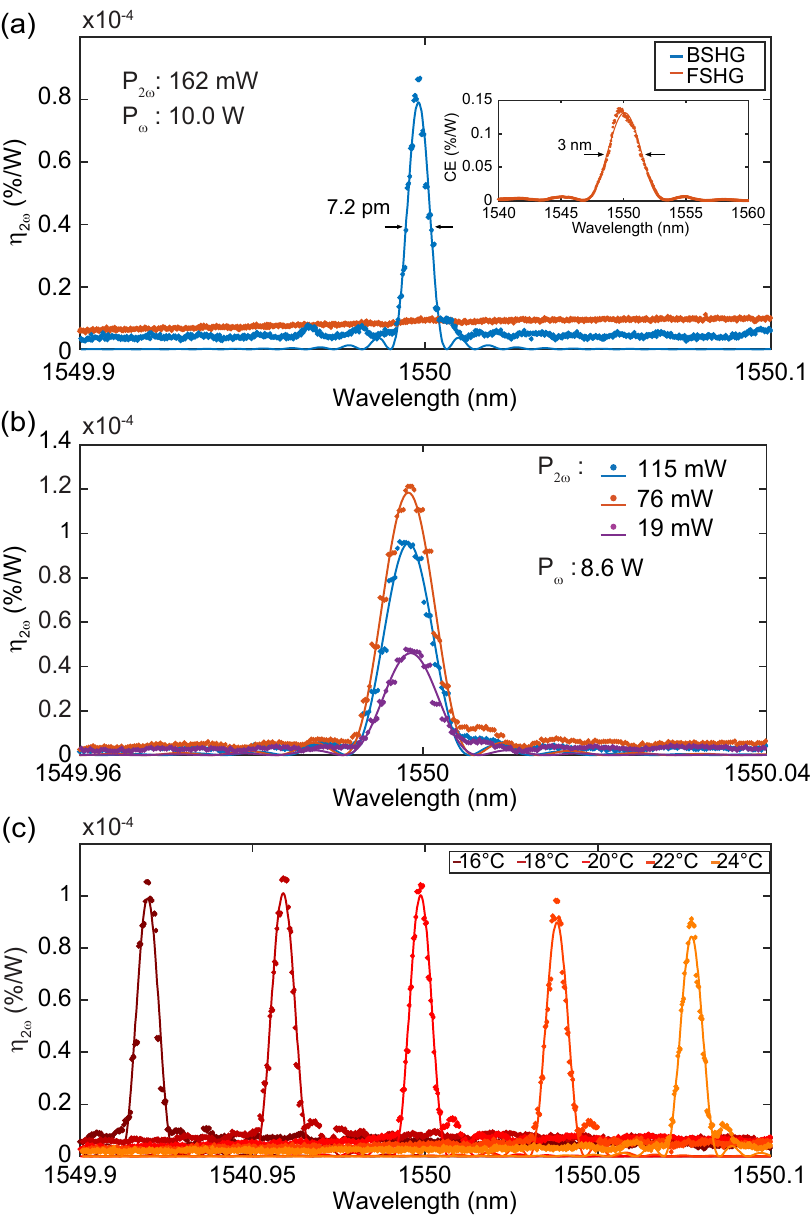}
\caption{a) Experimental BSHG CE spectrum (blue dots) and its sinc-squared fit using Eq. \eqref{ce} (blue solid line) FSHG CE spectrum shown in red for backward poling. Inset: FSHG CE spectrum of forward poling. b) CE spectra are shown for different SH seed powers. The steady state bandwidths are approximately identical for different seed powers. c) CE spectra (dots: experimental data, solid lines: sinc-squared fits) for temperatures from $16\,{\rm^\circ C}$ to $24\,{\rm ^\circ C}$. The waveguide was initially poled at 1550 nm and
$20\,{\rm^\circ C}$.
}
\label{fig:spectra}
\end{figure}

We measure the CE spectrum for both backward and forward directions. After AOP, if the propagation loss is ignored, the CE for BSHG becomes \cite{yakar2022coherent}: 
\begin{equation}
    \eta_{2\omega}\approx \frac{(\omega \chi^{(2)}_{\rm eff} L)^2}{2c^3\epsilon_0\bar{n}^2 \bar{S}} {{\rm sinc}^2\left( (\Delta k -\frac{2\pi}{\Lambda}){\tfrac{L}{2}}\right)}~,
    \label{ce}
\end{equation}
where $\Delta k(\lambda)-\frac{2\pi}{\Lambda}$ is the net wavevector mismatch after AOP, $L$ is the grating length, $\bar{n}=(n_{\omega}^2 n_{2\omega})^{1/3}$ with $n_{q\omega}$ being the effective refractive index at frequency $q\omega$ ($q=1,2$), $\bar{S}=(S_{\omega}^2 S_{2\omega})^{1/3}$ is the effective area, where $S_{q\omega}={\left(\iint dx dy |E_{q\omega}|^2\right)^{3/2}}/{\left(\iint dx dy |E_{q\omega}|^6\right)^{1/2}}$, $\chi^{(2)}_{\rm eff}$ is the effective second-order nonlinearity, and ${\rm sinc}(x)=\sin(x)/x$. The experimental data and the fits using Eq. \eqref{ce} are shown in Fig. \ref{fig:spectra}a. Here, the effective area and effective refractive index data used for fitting are obtained using finite element method simulations. We clearly see that forward SH is not generated while BSHG occurs with a very narrow bandwidth of $7.2\, {\rm pm}$ and a peak efficiency estimated around $1.2\times10^{-4}\,{\rm \%/W}$. This is in stark contrast to FSHG in the same waveguide after forward poling (see Fig. \ref{fig:setup}a inset) which results in a much broader bandwidth of $3\, {\rm nm}$. From the fit of the backward AOP CE spectrum, we extract the grating period, $\chi^{(2)}_{\rm eff}$ and grating length to be $206.3\, {\rm nm}$, $1.46\times10^{-3}\,{\rm pm/V}$ and $3.5\,{\rm cm}$ (81\% of the waveguide length), respectively. As seen from Fig. \ref{fig:spectra}b the steady state bandwidth is constant over different seeding conditions, which well agrees with Eq. \eqref{ce}.

The position of the CE peak can be thermally tuned \cite{fejer1992quasi, nitiss2020highly}, which for BSHG, becomes 
\begin{equation}
    \frac{\Delta \lambda}{\delta\lambda_{\rm FWHM}} \approx \frac{\frac{\partial n^t_{\rm eff}}{\partial T}Lc}{0.44\lambda_0^2} \Delta T ~,
\end{equation}
where $n^t_{\rm eff}=n_{\omega}+n_{2\omega}$, $\Delta T$ is the variation in temperature, $\Delta \lambda$ is the resulting detuning in wavelength and $\delta \lambda_{\rm FWHM}$ is the full-width half-maximum (FWHM) of QPM spectra. We experimentally investigate the thermal sensitivity of BSHG as shown in Fig. \ref{fig:spectra}c by varying the chip temperature. We extract the thermal shift of the QPM peak to be $19\,{\rm pm/^\circ C}$. It can be seen that it is extremely sensitive compared to its bandwidth.

In conclusion, seeded AOP represents a straightforward way to induce exceptionally short-period QPM gratings which can satisfy BSHG in integrated platforms. Seeded AOP circumvents fabrication challenges of the standard electrical poling techniques for implementation of sub-micron nonlinear gratings, while the self-organized QPM grating results in a near perfect match between experimental and theoretical spectral response. As predicted by the theory, we observe that only the nonlinear grating strength changes with changing seeding conditions. The poled waveguides show extremely narrow QPM spectra, highly tunable with temperature, with CE of $1.2\times10^{-4}\,{\rm \%/W}$, comparable to the highest one achieved in C-band so far \cite{busacca2014backward}. We theoretically explain the reduction of efficiencies compared to FSHG due to the screening of electric fields with sub-micron periods. The efficiency could be improved by using longer waveguides, exploiting other materials such as silicon-rich silicon nitride or using structures like Bragg gratings to separate the charges further.
\begin{acknowledgments}
This work was supported by ERC grant PISSARRO (ERC-2017-CoG 771647). The samples used for the experiment were fabricated by LIGENTEC SA. O.Y. would like to thank Dr. Christian Lafforgue and Mr. Serhat Bak{\i}rta\c{s} for valuable discussions. 
\end{acknowledgments}

\bibliography{apssamp}

\begin{thebibliography}{44}%
\makeatletter
\providecommand \@ifxundefined [1]{%
 \@ifx{#1\undefined}
}%
\providecommand \@ifnum [1]{%
 \ifnum #1\expandafter \@firstoftwo
 \else \expandafter \@secondoftwo
 \fi
}%
\providecommand \@ifx [1]{%
 \ifx #1\expandafter \@firstoftwo
 \else \expandafter \@secondoftwo
 \fi
}%
\providecommand \natexlab [1]{#1}%
\providecommand \enquote  [1]{``#1''}%
\providecommand \bibnamefont  [1]{#1}%
\providecommand \bibfnamefont [1]{#1}%
\providecommand \citenamefont [1]{#1}%
\providecommand \href@noop [0]{\@secondoftwo}%
\providecommand \href [0]{\begingroup \@sanitize@url \@href}%
\providecommand \@href[1]{\@@startlink{#1}\@@href}%
\providecommand \@@href[1]{\endgroup#1\@@endlink}%
\providecommand \@sanitize@url [0]{\catcode `\\12\catcode `\$12\catcode
  `\&12\catcode `\#12\catcode `\^12\catcode `\_12\catcode `\%12\relax}%
\providecommand \@@startlink[1]{}%
\providecommand \@@endlink[0]{}%
\providecommand \url  [0]{\begingroup\@sanitize@url \@url }%
\providecommand \@url [1]{\endgroup\@href {#1}{\urlprefix }}%
\providecommand \urlprefix  [0]{URL }%
\providecommand \Eprint [0]{\href }%
\providecommand \doibase [0]{https://doi.org/}%
\providecommand \selectlanguage [0]{\@gobble}%
\providecommand \bibinfo  [0]{\@secondoftwo}%
\providecommand \bibfield  [0]{\@secondoftwo}%
\providecommand \translation [1]{[#1]}%
\providecommand \BibitemOpen [0]{}%
\providecommand \bibitemStop [0]{}%
\providecommand \bibitemNoStop [0]{.\EOS\space}%
\providecommand \EOS [0]{\spacefactor3000\relax}%
\providecommand \BibitemShut  [1]{\csname bibitem#1\endcsname}%
\let\auto@bib@innerbib\@empty
\bibitem [{\citenamefont {Armstrong}\ \emph {et~al.}(1962)\citenamefont
  {Armstrong}, \citenamefont {Bloembergen}, \citenamefont {Ducuing},\ and\
  \citenamefont {Pershan}}]{armstrong1962interactions}%
  \BibitemOpen
  \bibfield  {author} {\bibinfo {author} {\bibfnamefont {J.~A.}\ \bibnamefont
  {Armstrong}}, \bibinfo {author} {\bibfnamefont {N.}~\bibnamefont
  {Bloembergen}}, \bibinfo {author} {\bibfnamefont {J.}~\bibnamefont
  {Ducuing}},\ and\ \bibinfo {author} {\bibfnamefont {P.~S.}\ \bibnamefont
  {Pershan}},\ }\bibfield  {title} {\bibinfo {title} {Interactions between
  light waves in a nonlinear dielectric},\ }\href
  {https://doi.org/10.1103/PhysRev.127.1918} {\bibfield  {journal} {\bibinfo
  {journal} {Phys. Rev.}\ }\textbf {\bibinfo {volume} {127}},\ \bibinfo {pages}
  {1918} (\bibinfo {year} {1962})}\BibitemShut {NoStop}%
\bibitem [{\citenamefont {Wang}\ \emph {et~al.}(2018)\citenamefont {Wang},
  \citenamefont {Langrock}, \citenamefont {Marandi}, \citenamefont {Jankowski},
  \citenamefont {Zhang}, \citenamefont {Desiatov}, \citenamefont {Fejer},\ and\
  \citenamefont {Lon{\v{c}}ar}}]{wang2018ultrahigh}%
  \BibitemOpen
  \bibfield  {author} {\bibinfo {author} {\bibfnamefont {C.}~\bibnamefont
  {Wang}}, \bibinfo {author} {\bibfnamefont {C.}~\bibnamefont {Langrock}},
  \bibinfo {author} {\bibfnamefont {A.}~\bibnamefont {Marandi}}, \bibinfo
  {author} {\bibfnamefont {M.}~\bibnamefont {Jankowski}}, \bibinfo {author}
  {\bibfnamefont {M.}~\bibnamefont {Zhang}}, \bibinfo {author} {\bibfnamefont
  {B.}~\bibnamefont {Desiatov}}, \bibinfo {author} {\bibfnamefont {M.~M.}\
  \bibnamefont {Fejer}},\ and\ \bibinfo {author} {\bibfnamefont
  {M.}~\bibnamefont {Lon{\v{c}}ar}},\ }\bibfield  {title} {\bibinfo {title}
  {Ultrahigh-efficiency wavelength conversion in nanophotonic periodically
  poled lithium niobate waveguides},\ }\href@noop {} {\bibfield  {journal}
  {\bibinfo  {journal} {Optica}\ }\textbf {\bibinfo {volume} {5}},\ \bibinfo
  {pages} {1438} (\bibinfo {year} {2018})}\BibitemShut {NoStop}%
\bibitem [{\citenamefont {Timurdogan}\ \emph {et~al.}(2017)\citenamefont
  {Timurdogan}, \citenamefont {Poulton}, \citenamefont {Byrd},\ and\
  \citenamefont {Watts}}]{timurdogan2017electric}%
  \BibitemOpen
  \bibfield  {author} {\bibinfo {author} {\bibfnamefont {E.}~\bibnamefont
  {Timurdogan}}, \bibinfo {author} {\bibfnamefont {C.~V.}\ \bibnamefont
  {Poulton}}, \bibinfo {author} {\bibfnamefont {M.}~\bibnamefont {Byrd}},\ and\
  \bibinfo {author} {\bibfnamefont {M.}~\bibnamefont {Watts}},\ }\bibfield
  {title} {\bibinfo {title} {Electric field-induced second-order nonlinear
  optical effects in silicon waveguides},\ }\href@noop {} {\bibfield  {journal}
  {\bibinfo  {journal} {Nature Photonics}\ }\textbf {\bibinfo {volume} {11}},\
  \bibinfo {pages} {200} (\bibinfo {year} {2017})}\BibitemShut {NoStop}%
\bibitem [{\citenamefont {{\"O}sterberg}\ and\ \citenamefont
  {Margulis}(1986)}]{osterberg1986dye}%
  \BibitemOpen
  \bibfield  {author} {\bibinfo {author} {\bibfnamefont {U.}~\bibnamefont
  {{\"O}sterberg}}\ and\ \bibinfo {author} {\bibfnamefont {W.}~\bibnamefont
  {Margulis}},\ }\bibfield  {title} {\bibinfo {title} {Dye laser pumped by nd:
  Yag laser pulses frequency doubled in a glass optical fiber},\ }\href@noop {}
  {\bibfield  {journal} {\bibinfo  {journal} {Optics letters}\ }\textbf
  {\bibinfo {volume} {11}},\ \bibinfo {pages} {516} (\bibinfo {year}
  {1986})}\BibitemShut {NoStop}%
\bibitem [{\citenamefont {Baskin}\ and\ \citenamefont
  {Entin}(1988)}]{baskin1988coherent}%
  \BibitemOpen
  \bibfield  {author} {\bibinfo {author} {\bibfnamefont {E.}~\bibnamefont
  {Baskin}}\ and\ \bibinfo {author} {\bibfnamefont {M.}~\bibnamefont {Entin}},\
  }\bibfield  {title} {\bibinfo {title} {Coherent photovoltaic effect due to
  the quantum corrections},\ }\href@noop {} {\bibfield  {journal} {\bibinfo
  {journal} {Soviet Journal of Experimental and Theoretical Physics Letters}\
  }\textbf {\bibinfo {volume} {48}},\ \bibinfo {pages} {601} (\bibinfo {year}
  {1988})}\BibitemShut {NoStop}%
\bibitem [{\citenamefont {Zel'Dovich}\ and\ \citenamefont
  {Chudinov}(1989)}]{zel1989interference}%
  \BibitemOpen
  \bibfield  {author} {\bibinfo {author} {\bibfnamefont {B.~Y.}\ \bibnamefont
  {Zel'Dovich}}\ and\ \bibinfo {author} {\bibfnamefont {A.}~\bibnamefont
  {Chudinov}},\ }\bibfield  {title} {\bibinfo {title} {Interference of fields
  with frequencies $\omega$ and 2$\omega$ in external photoelectric effect},\
  }\href@noop {} {\bibfield  {journal} {\bibinfo  {journal} {Soviet Journal of
  Experimental and Theoretical Physics Letters}\ }\textbf {\bibinfo {volume}
  {50}},\ \bibinfo {pages} {439} (\bibinfo {year} {1989})}\BibitemShut
  {NoStop}%
\bibitem [{\citenamefont {Anderson}\ \emph {et~al.}(1991)\citenamefont
  {Anderson}, \citenamefont {Mizrahi},\ and\ \citenamefont
  {Sipe}}]{sipeol1991}%
  \BibitemOpen
  \bibfield  {author} {\bibinfo {author} {\bibfnamefont {D.~Z.}\ \bibnamefont
  {Anderson}}, \bibinfo {author} {\bibfnamefont {V.}~\bibnamefont {Mizrahi}},\
  and\ \bibinfo {author} {\bibfnamefont {J.~E.}\ \bibnamefont {Sipe}},\
  }\bibfield  {title} {\bibinfo {title} {Model for second-harmonic generation
  in glass optical fibers based on asymmetric photoelectron emission from
  defect sites},\ }\href@noop {} {\bibfield  {journal} {\bibinfo  {journal}
  {Optics letters}\ }\textbf {\bibinfo {volume} {16}},\ \bibinfo {pages} {796}
  (\bibinfo {year} {1991})}\BibitemShut {NoStop}%
\bibitem [{\citenamefont {Margulis}\ \emph {et~al.}(1995)\citenamefont
  {Margulis}, \citenamefont {Laurell},\ and\ \citenamefont
  {Lesche}}]{margulis1995imaging}%
  \BibitemOpen
  \bibfield  {author} {\bibinfo {author} {\bibfnamefont {W.}~\bibnamefont
  {Margulis}}, \bibinfo {author} {\bibfnamefont {F.}~\bibnamefont {Laurell}},\
  and\ \bibinfo {author} {\bibfnamefont {B.}~\bibnamefont {Lesche}},\
  }\bibfield  {title} {\bibinfo {title} {Imaging the nonlinear grating in
  frequency-doubling fibres},\ }\href@noop {} {\bibfield  {journal} {\bibinfo
  {journal} {Nature}\ }\textbf {\bibinfo {volume} {378}},\ \bibinfo {pages}
  {699} (\bibinfo {year} {1995})}\BibitemShut {NoStop}%
\bibitem [{\citenamefont {Dianov}\ and\ \citenamefont
  {Starodubov}(1995)}]{dianov1995photoinduced}%
  \BibitemOpen
  \bibfield  {author} {\bibinfo {author} {\bibfnamefont {E.~M.}\ \bibnamefont
  {Dianov}}\ and\ \bibinfo {author} {\bibfnamefont {D.~S.}\ \bibnamefont
  {Starodubov}},\ }\bibfield  {title} {\bibinfo {title} {Photoinduced
  generation of the second harmonic in centrosymmetric media},\ }\href@noop {}
  {\bibfield  {journal} {\bibinfo  {journal} {Quantum Electronics}\ }\textbf
  {\bibinfo {volume} {25}},\ \bibinfo {pages} {395} (\bibinfo {year}
  {1995})}\BibitemShut {NoStop}%
\bibitem [{\citenamefont {Billat}\ \emph {et~al.}(2017)\citenamefont {Billat},
  \citenamefont {Grassani}, \citenamefont {Pfeiffer}, \citenamefont
  {Kharitonov}, \citenamefont {Kippenberg},\ and\ \citenamefont
  {Br{\`e}s}}]{billat2017large}%
  \BibitemOpen
  \bibfield  {author} {\bibinfo {author} {\bibfnamefont {A.}~\bibnamefont
  {Billat}}, \bibinfo {author} {\bibfnamefont {D.}~\bibnamefont {Grassani}},
  \bibinfo {author} {\bibfnamefont {M.~H.}\ \bibnamefont {Pfeiffer}}, \bibinfo
  {author} {\bibfnamefont {S.}~\bibnamefont {Kharitonov}}, \bibinfo {author}
  {\bibfnamefont {T.~J.}\ \bibnamefont {Kippenberg}},\ and\ \bibinfo {author}
  {\bibfnamefont {C.-S.}\ \bibnamefont {Br{\`e}s}},\ }\bibfield  {title}
  {\bibinfo {title} {Large second harmonic generation enhancement in
  {S}i$_3${N}$_4$ waveguides by all-optically induced quasi-phase-matching},\
  }\href@noop {} {\bibfield  {journal} {\bibinfo  {journal} {Nature
  communications}\ }\textbf {\bibinfo {volume} {8}},\ \bibinfo {pages} {1}
  (\bibinfo {year} {2017})}\BibitemShut {NoStop}%
\bibitem [{\citenamefont {Porcel}\ \emph {et~al.}(2017)\citenamefont {Porcel},
  \citenamefont {Mak}, \citenamefont {Taballione}, \citenamefont
  {Schermerhorn}, \citenamefont {Epping}, \citenamefont {van~der Slot},\ and\
  \citenamefont {Boller}}]{porcel2017photo}%
  \BibitemOpen
  \bibfield  {author} {\bibinfo {author} {\bibfnamefont {M.~A.}\ \bibnamefont
  {Porcel}}, \bibinfo {author} {\bibfnamefont {J.}~\bibnamefont {Mak}},
  \bibinfo {author} {\bibfnamefont {C.}~\bibnamefont {Taballione}}, \bibinfo
  {author} {\bibfnamefont {V.~K.}\ \bibnamefont {Schermerhorn}}, \bibinfo
  {author} {\bibfnamefont {J.~P.}\ \bibnamefont {Epping}}, \bibinfo {author}
  {\bibfnamefont {P.~J.}\ \bibnamefont {van~der Slot}},\ and\ \bibinfo {author}
  {\bibfnamefont {K.-J.}\ \bibnamefont {Boller}},\ }\bibfield  {title}
  {\bibinfo {title} {Photo-induced second-order nonlinearity in stoichiometric
  silicon nitride waveguides},\ }\href@noop {} {\bibfield  {journal} {\bibinfo
  {journal} {Optics express}\ }\textbf {\bibinfo {volume} {25}},\ \bibinfo
  {pages} {33143} (\bibinfo {year} {2017})}\BibitemShut {NoStop}%
\bibitem [{\citenamefont {Hickstein}\ \emph {et~al.}(2019)\citenamefont
  {Hickstein}, \citenamefont {Carlson}, \citenamefont {Mundoor}, \citenamefont
  {Khurgin}, \citenamefont {Srinivasan}, \citenamefont {Westly}, \citenamefont
  {Kowligy}, \citenamefont {Smalyukh}, \citenamefont {Diddams},\ and\
  \citenamefont {Papp}}]{hickstein2019self}%
  \BibitemOpen
  \bibfield  {author} {\bibinfo {author} {\bibfnamefont {D.~D.}\ \bibnamefont
  {Hickstein}}, \bibinfo {author} {\bibfnamefont {D.~R.}\ \bibnamefont
  {Carlson}}, \bibinfo {author} {\bibfnamefont {H.}~\bibnamefont {Mundoor}},
  \bibinfo {author} {\bibfnamefont {J.~B.}\ \bibnamefont {Khurgin}}, \bibinfo
  {author} {\bibfnamefont {K.}~\bibnamefont {Srinivasan}}, \bibinfo {author}
  {\bibfnamefont {D.}~\bibnamefont {Westly}}, \bibinfo {author} {\bibfnamefont
  {A.}~\bibnamefont {Kowligy}}, \bibinfo {author} {\bibfnamefont {I.~I.}\
  \bibnamefont {Smalyukh}}, \bibinfo {author} {\bibfnamefont {S.~A.}\
  \bibnamefont {Diddams}},\ and\ \bibinfo {author} {\bibfnamefont {S.~B.}\
  \bibnamefont {Papp}},\ }\bibfield  {title} {\bibinfo {title} {Self-organized
  nonlinear gratings for ultrafast nanophotonics},\ }\href@noop {} {\bibfield
  {journal} {\bibinfo  {journal} {Nature Photonics}\ }\textbf {\bibinfo
  {volume} {13}},\ \bibinfo {pages} {494} (\bibinfo {year} {2019})}\BibitemShut
  {NoStop}%
\bibitem [{\citenamefont {Lu}\ \emph {et~al.}(2021)\citenamefont {Lu},
  \citenamefont {Moille}, \citenamefont {Rao}, \citenamefont {Westly},\ and\
  \citenamefont {Srinivasan}}]{lu2021efficient}%
  \BibitemOpen
  \bibfield  {author} {\bibinfo {author} {\bibfnamefont {X.}~\bibnamefont
  {Lu}}, \bibinfo {author} {\bibfnamefont {G.}~\bibnamefont {Moille}}, \bibinfo
  {author} {\bibfnamefont {A.}~\bibnamefont {Rao}}, \bibinfo {author}
  {\bibfnamefont {D.~A.}\ \bibnamefont {Westly}},\ and\ \bibinfo {author}
  {\bibfnamefont {K.}~\bibnamefont {Srinivasan}},\ }\bibfield  {title}
  {\bibinfo {title} {Efficient photoinduced second-harmonic generation in
  silicon nitride photonics},\ }\href@noop {} {\bibfield  {journal} {\bibinfo
  {journal} {Nature Photonics}\ }\textbf {\bibinfo {volume} {15}},\ \bibinfo
  {pages} {131} (\bibinfo {year} {2021})}\BibitemShut {NoStop}%
\bibitem [{\citenamefont {Nitiss}\ \emph {et~al.}(2022)\citenamefont {Nitiss},
  \citenamefont {Hu}, \citenamefont {Stroganov},\ and\ \citenamefont
  {Brès}}]{nitiss2021optically}%
  \BibitemOpen
  \bibfield  {author} {\bibinfo {author} {\bibfnamefont {E.}~\bibnamefont
  {Nitiss}}, \bibinfo {author} {\bibfnamefont {J.}~\bibnamefont {Hu}}, \bibinfo
  {author} {\bibfnamefont {A.}~\bibnamefont {Stroganov}},\ and\ \bibinfo
  {author} {\bibfnamefont {C.-S.}\ \bibnamefont {Brès}},\ }\bibfield  {title}
  {\bibinfo {title} {Optically reconfigurable quasi-phase-matching in silicon
  nitride microresonators},\ }\bibfield  {journal} {\bibinfo  {journal} {Nature
  Photonics}\ }\href {https://doi.org/10.1038/s41566-021-00925-5}
  {10.1038/s41566-021-00925-5} (\bibinfo {year} {2022})\BibitemShut {NoStop}%
\bibitem [{\citenamefont {Hu}\ \emph {et~al.}(2022)\citenamefont {Hu},
  \citenamefont {Nitiss}, \citenamefont {He}, \citenamefont {Liu},
  \citenamefont {Yakar}, \citenamefont {Weng}, \citenamefont {Kippenberg},\
  and\ \citenamefont {Br{\`e}s}}]{hu2022photo}%
  \BibitemOpen
  \bibfield  {author} {\bibinfo {author} {\bibfnamefont {J.}~\bibnamefont
  {Hu}}, \bibinfo {author} {\bibfnamefont {E.}~\bibnamefont {Nitiss}}, \bibinfo
  {author} {\bibfnamefont {J.}~\bibnamefont {He}}, \bibinfo {author}
  {\bibfnamefont {J.}~\bibnamefont {Liu}}, \bibinfo {author} {\bibfnamefont
  {O.}~\bibnamefont {Yakar}}, \bibinfo {author} {\bibfnamefont
  {W.}~\bibnamefont {Weng}}, \bibinfo {author} {\bibfnamefont {T.~J.}\
  \bibnamefont {Kippenberg}},\ and\ \bibinfo {author} {\bibfnamefont {C.-S.}\
  \bibnamefont {Br{\`e}s}},\ }\bibfield  {title} {\bibinfo {title}
  {Photo-induced cascaded harmonic and comb generation in silicon nitride
  microresonators},\ }\href@noop {} {\bibfield  {journal} {\bibinfo  {journal}
  {Science Advances}\ }\textbf {\bibinfo {volume} {8}},\ \bibinfo {pages}
  {eadd8252} (\bibinfo {year} {2022})}\BibitemShut {NoStop}%
\bibitem [{\citenamefont {Ye}\ \emph {et~al.}(2021)\citenamefont {Ye},
  \citenamefont {Zhao}, \citenamefont {Twayana}, \citenamefont {Karlsson},
  \citenamefont {Torres-Company},\ and\ \citenamefont
  {Andrekson}}]{ye2021overcoming}%
  \BibitemOpen
  \bibfield  {author} {\bibinfo {author} {\bibfnamefont {Z.}~\bibnamefont
  {Ye}}, \bibinfo {author} {\bibfnamefont {P.}~\bibnamefont {Zhao}}, \bibinfo
  {author} {\bibfnamefont {K.}~\bibnamefont {Twayana}}, \bibinfo {author}
  {\bibfnamefont {M.}~\bibnamefont {Karlsson}}, \bibinfo {author}
  {\bibfnamefont {V.}~\bibnamefont {Torres-Company}},\ and\ \bibinfo {author}
  {\bibfnamefont {P.~A.}\ \bibnamefont {Andrekson}},\ }\bibfield  {title}
  {\bibinfo {title} {Overcoming the quantum limit of optical amplification in
  monolithic waveguides},\ }\href@noop {} {\bibfield  {journal} {\bibinfo
  {journal} {Science advances}\ }\textbf {\bibinfo {volume} {7}},\ \bibinfo
  {pages} {eabi8150} (\bibinfo {year} {2021})}\BibitemShut {NoStop}%
\bibitem [{\citenamefont {Ayan}\ \emph {et~al.}(2022)\citenamefont {Ayan},
  \citenamefont {Mazeas}, \citenamefont {Liu}, \citenamefont {Kippenberg},\
  and\ \citenamefont {Br\`{e}s}}]{ayan2022polarization}%
  \BibitemOpen
  \bibfield  {author} {\bibinfo {author} {\bibfnamefont {A.}~\bibnamefont
  {Ayan}}, \bibinfo {author} {\bibfnamefont {F.}~\bibnamefont {Mazeas}},
  \bibinfo {author} {\bibfnamefont {J.}~\bibnamefont {Liu}}, \bibinfo {author}
  {\bibfnamefont {T.~J.}\ \bibnamefont {Kippenberg}},\ and\ \bibinfo {author}
  {\bibfnamefont {C.-S.}\ \bibnamefont {Br\`{e}s}},\ }\bibfield  {title}
  {\bibinfo {title} {Polarization selective ultra-broadband wavelength
  conversion in silicon nitride waveguides},\ }\href
  {https://doi.org/10.1364/OE.446357} {\bibfield  {journal} {\bibinfo
  {journal} {Opt. Express}\ }\textbf {\bibinfo {volume} {30}},\ \bibinfo
  {pages} {4342} (\bibinfo {year} {2022})}\BibitemShut {NoStop}%
\bibitem [{\citenamefont {Levy}\ \emph {et~al.}(2011)\citenamefont {Levy},
  \citenamefont {Foster}, \citenamefont {Gaeta},\ and\ \citenamefont
  {Lipson}}]{levy2011harmonic}%
  \BibitemOpen
  \bibfield  {author} {\bibinfo {author} {\bibfnamefont {J.~S.}\ \bibnamefont
  {Levy}}, \bibinfo {author} {\bibfnamefont {M.~A.}\ \bibnamefont {Foster}},
  \bibinfo {author} {\bibfnamefont {A.~L.}\ \bibnamefont {Gaeta}},\ and\
  \bibinfo {author} {\bibfnamefont {M.}~\bibnamefont {Lipson}},\ }\bibfield
  {title} {\bibinfo {title} {Harmonic generation in silicon nitride ring
  resonators},\ }\href@noop {} {\bibfield  {journal} {\bibinfo  {journal}
  {Optics express}\ }\textbf {\bibinfo {volume} {19}},\ \bibinfo {pages}
  {11415} (\bibinfo {year} {2011})}\BibitemShut {NoStop}%
\bibitem [{\citenamefont {Grassani}\ \emph {et~al.}(2019)\citenamefont
  {Grassani}, \citenamefont {Tagkoudi}, \citenamefont {Guo}, \citenamefont
  {Herkommer}, \citenamefont {Yang}, \citenamefont {Kippenberg},\ and\
  \citenamefont {Br{\`e}s}}]{grassani2019mid}%
  \BibitemOpen
  \bibfield  {author} {\bibinfo {author} {\bibfnamefont {D.}~\bibnamefont
  {Grassani}}, \bibinfo {author} {\bibfnamefont {E.}~\bibnamefont {Tagkoudi}},
  \bibinfo {author} {\bibfnamefont {H.}~\bibnamefont {Guo}}, \bibinfo {author}
  {\bibfnamefont {C.}~\bibnamefont {Herkommer}}, \bibinfo {author}
  {\bibfnamefont {F.}~\bibnamefont {Yang}}, \bibinfo {author} {\bibfnamefont
  {T.~J.}\ \bibnamefont {Kippenberg}},\ and\ \bibinfo {author} {\bibfnamefont
  {C.-S.}\ \bibnamefont {Br{\`e}s}},\ }\bibfield  {title} {\bibinfo {title}
  {Mid infrared gas spectroscopy using efficient fiber laser driven photonic
  chip-based supercontinuum},\ }\href@noop {} {\bibfield  {journal} {\bibinfo
  {journal} {Nature communications}\ }\textbf {\bibinfo {volume} {10}},\
  \bibinfo {pages} {1} (\bibinfo {year} {2019})}\BibitemShut {NoStop}%
\bibitem [{\citenamefont {Kippenberg}\ \emph {et~al.}(2018)\citenamefont
  {Kippenberg}, \citenamefont {Gaeta}, \citenamefont {Lipson},\ and\
  \citenamefont {Gorodetsky}}]{kippenberg2018dissipative}%
  \BibitemOpen
  \bibfield  {author} {\bibinfo {author} {\bibfnamefont {T.~J.}\ \bibnamefont
  {Kippenberg}}, \bibinfo {author} {\bibfnamefont {A.~L.}\ \bibnamefont
  {Gaeta}}, \bibinfo {author} {\bibfnamefont {M.}~\bibnamefont {Lipson}},\ and\
  \bibinfo {author} {\bibfnamefont {M.~L.}\ \bibnamefont {Gorodetsky}},\
  }\bibfield  {title} {\bibinfo {title} {Dissipative kerr solitons in optical
  microresonators},\ }\href@noop {} {\bibfield  {journal} {\bibinfo  {journal}
  {Science}\ }\textbf {\bibinfo {volume} {361}} (\bibinfo {year}
  {2018})}\BibitemShut {NoStop}%
\bibitem [{\citenamefont {Nitiss}\ \emph
  {et~al.}(2020{\natexlab{a}})\citenamefont {Nitiss}, \citenamefont {Zabelich},
  \citenamefont {Yakar}, \citenamefont {Liu}, \citenamefont {Wang},
  \citenamefont {Kippenberg},\ and\ \citenamefont
  {Br{\`e}s}}]{nitiss2020broadband}%
  \BibitemOpen
  \bibfield  {author} {\bibinfo {author} {\bibfnamefont {E.}~\bibnamefont
  {Nitiss}}, \bibinfo {author} {\bibfnamefont {B.}~\bibnamefont {Zabelich}},
  \bibinfo {author} {\bibfnamefont {O.}~\bibnamefont {Yakar}}, \bibinfo
  {author} {\bibfnamefont {J.}~\bibnamefont {Liu}}, \bibinfo {author}
  {\bibfnamefont {R.~N.}\ \bibnamefont {Wang}}, \bibinfo {author}
  {\bibfnamefont {T.~J.}\ \bibnamefont {Kippenberg}},\ and\ \bibinfo {author}
  {\bibfnamefont {C.-S.}\ \bibnamefont {Br{\`e}s}},\ }\bibfield  {title}
  {\bibinfo {title} {Broadband quasi-phase-matching in dispersion-engineered
  all-optically poled silicon nitride waveguides},\ }\href@noop {} {\bibfield
  {journal} {\bibinfo  {journal} {Photonics Research}\ }\textbf {\bibinfo
  {volume} {8}},\ \bibinfo {pages} {1475} (\bibinfo {year}
  {2020}{\natexlab{a}})}\BibitemShut {NoStop}%
\bibitem [{\citenamefont {Sahin}\ \emph {et~al.}(2021)\citenamefont {Sahin},
  \citenamefont {Zabelich}, \citenamefont {Yakar}, \citenamefont {Nitiss},
  \citenamefont {Liu}, \citenamefont {Wang}, \citenamefont {Kippenberg},\ and\
  \citenamefont {Br{\`e}s}}]{sahin2021difference}%
  \BibitemOpen
  \bibfield  {author} {\bibinfo {author} {\bibfnamefont {E.}~\bibnamefont
  {Sahin}}, \bibinfo {author} {\bibfnamefont {B.}~\bibnamefont {Zabelich}},
  \bibinfo {author} {\bibfnamefont {O.}~\bibnamefont {Yakar}}, \bibinfo
  {author} {\bibfnamefont {E.}~\bibnamefont {Nitiss}}, \bibinfo {author}
  {\bibfnamefont {J.}~\bibnamefont {Liu}}, \bibinfo {author} {\bibfnamefont
  {R.~N.}\ \bibnamefont {Wang}}, \bibinfo {author} {\bibfnamefont {T.~J.}\
  \bibnamefont {Kippenberg}},\ and\ \bibinfo {author} {\bibfnamefont {C.-S.}\
  \bibnamefont {Br{\`e}s}},\ }\bibfield  {title} {\bibinfo {title}
  {Difference-frequency generation in optically poled silicon nitride
  waveguides},\ }\href@noop {} {\bibfield  {journal} {\bibinfo  {journal}
  {Nanophotonics}\ }\textbf {\bibinfo {volume} {10}},\ \bibinfo {pages} {1923}
  (\bibinfo {year} {2021})}\BibitemShut {NoStop}%
\bibitem [{\citenamefont {Dalidet}\ \emph {et~al.}(2022)\citenamefont
  {Dalidet}, \citenamefont {Mazeas}, \citenamefont {Nitiss}, \citenamefont
  {Yakar}, \citenamefont {Stroganov}, \citenamefont {Tanzilli}, \citenamefont
  {Labont{\'e}},\ and\ \citenamefont {Br{\`e}s}}]{dalidet2022near}%
  \BibitemOpen
  \bibfield  {author} {\bibinfo {author} {\bibfnamefont {R.}~\bibnamefont
  {Dalidet}}, \bibinfo {author} {\bibfnamefont {F.}~\bibnamefont {Mazeas}},
  \bibinfo {author} {\bibfnamefont {E.}~\bibnamefont {Nitiss}}, \bibinfo
  {author} {\bibfnamefont {O.}~\bibnamefont {Yakar}}, \bibinfo {author}
  {\bibfnamefont {A.}~\bibnamefont {Stroganov}}, \bibinfo {author}
  {\bibfnamefont {S.}~\bibnamefont {Tanzilli}}, \bibinfo {author}
  {\bibfnamefont {L.}~\bibnamefont {Labont{\'e}}},\ and\ \bibinfo {author}
  {\bibfnamefont {C.-S.}\ \bibnamefont {Br{\`e}s}},\ }\bibfield  {title}
  {\bibinfo {title} {Near perfect two-photon interference out of a
  down-converter on a silicon photonic chip},\ }\href@noop {} {\bibfield
  {journal} {\bibinfo  {journal} {Optics Express}\ }\textbf {\bibinfo {volume}
  {30}},\ \bibinfo {pages} {11298} (\bibinfo {year} {2022})}\BibitemShut
  {NoStop}%
\bibitem [{\citenamefont {Canalias}\ and\ \citenamefont
  {Pasiskevicius}(2007)}]{canalias2007mirrorless}%
  \BibitemOpen
  \bibfield  {author} {\bibinfo {author} {\bibfnamefont {C.}~\bibnamefont
  {Canalias}}\ and\ \bibinfo {author} {\bibfnamefont {V.}~\bibnamefont
  {Pasiskevicius}},\ }\bibfield  {title} {\bibinfo {title} {Mirrorless optical
  parametric oscillator},\ }\href@noop {} {\bibfield  {journal} {\bibinfo
  {journal} {Nature Photonics}\ }\textbf {\bibinfo {volume} {1}},\ \bibinfo
  {pages} {459} (\bibinfo {year} {2007})}\BibitemShut {NoStop}%
\bibitem [{\citenamefont {Luo}\ \emph {et~al.}(2020)\citenamefont {Luo},
  \citenamefont {Ansari}, \citenamefont {Massaro}, \citenamefont {Santandrea},
  \citenamefont {Eigner}, \citenamefont {Ricken}, \citenamefont {Herrmann},\
  and\ \citenamefont {Silberhorn}}]{luo2020counter}%
  \BibitemOpen
  \bibfield  {author} {\bibinfo {author} {\bibfnamefont {K.-H.}\ \bibnamefont
  {Luo}}, \bibinfo {author} {\bibfnamefont {V.}~\bibnamefont {Ansari}},
  \bibinfo {author} {\bibfnamefont {M.}~\bibnamefont {Massaro}}, \bibinfo
  {author} {\bibfnamefont {M.}~\bibnamefont {Santandrea}}, \bibinfo {author}
  {\bibfnamefont {C.}~\bibnamefont {Eigner}}, \bibinfo {author} {\bibfnamefont
  {R.}~\bibnamefont {Ricken}}, \bibinfo {author} {\bibfnamefont
  {H.}~\bibnamefont {Herrmann}},\ and\ \bibinfo {author} {\bibfnamefont
  {C.}~\bibnamefont {Silberhorn}},\ }\bibfield  {title} {\bibinfo {title}
  {Counter-propagating photon pair generation in a nonlinear waveguide},\
  }\href {https://doi.org/10.1364/OE.378789} {\bibfield  {journal} {\bibinfo
  {journal} {Opt. Express}\ }\textbf {\bibinfo {volume} {28}},\ \bibinfo
  {pages} {3215} (\bibinfo {year} {2020})}\BibitemShut {NoStop}%
\bibitem [{\citenamefont {Liu}\ \emph {et~al.}(2021)\citenamefont {Liu},
  \citenamefont {Guo}, \citenamefont {Ren}, \citenamefont {Yang}, \citenamefont
  {Shang}, \citenamefont {Zhou}, \citenamefont {Li}, \citenamefont {Sun},
  \citenamefont {Xu}, \citenamefont {Xie} \emph {et~al.}}]{liu2021observation}%
  \BibitemOpen
  \bibfield  {author} {\bibinfo {author} {\bibfnamefont {Y.-C.}\ \bibnamefont
  {Liu}}, \bibinfo {author} {\bibfnamefont {D.-J.}\ \bibnamefont {Guo}},
  \bibinfo {author} {\bibfnamefont {K.-Q.}\ \bibnamefont {Ren}}, \bibinfo
  {author} {\bibfnamefont {R.}~\bibnamefont {Yang}}, \bibinfo {author}
  {\bibfnamefont {M.}~\bibnamefont {Shang}}, \bibinfo {author} {\bibfnamefont
  {W.}~\bibnamefont {Zhou}}, \bibinfo {author} {\bibfnamefont {X.}~\bibnamefont
  {Li}}, \bibinfo {author} {\bibfnamefont {C.-W.}\ \bibnamefont {Sun}},
  \bibinfo {author} {\bibfnamefont {P.}~\bibnamefont {Xu}}, \bibinfo {author}
  {\bibfnamefont {Z.}~\bibnamefont {Xie}}, \emph {et~al.},\ }\bibfield  {title}
  {\bibinfo {title} {Observation of frequency-uncorrelated photon pairs
  generated by counter-propagating spontaneous parametric down-conversion},\
  }\href@noop {} {\bibfield  {journal} {\bibinfo  {journal} {Scientific
  Reports}\ }\textbf {\bibinfo {volume} {11}},\ \bibinfo {pages} {1} (\bibinfo
  {year} {2021})}\BibitemShut {NoStop}%
\bibitem [{\citenamefont {D'Alessandro}\ \emph {et~al.}(1997)\citenamefont
  {D'Alessandro}, \citenamefont {Russell},\ and\ \citenamefont
  {Wheeler}}]{dalessandro1997nonlinear}%
  \BibitemOpen
  \bibfield  {author} {\bibinfo {author} {\bibfnamefont {G.}~\bibnamefont
  {D'Alessandro}}, \bibinfo {author} {\bibfnamefont {P.~S.~J.}\ \bibnamefont
  {Russell}},\ and\ \bibinfo {author} {\bibfnamefont {A.~A.}\ \bibnamefont
  {Wheeler}},\ }\bibfield  {title} {\bibinfo {title} {Nonlinear dynamics of a
  backward quasi-phase-matched second-harmonic generator},\ }\href
  {https://doi.org/10.1103/PhysRevA.55.3211} {\bibfield  {journal} {\bibinfo
  {journal} {Phys. Rev. A}\ }\textbf {\bibinfo {volume} {55}},\ \bibinfo
  {pages} {3211} (\bibinfo {year} {1997})}\BibitemShut {NoStop}%
\bibitem [{\citenamefont {Boes}\ \emph {et~al.}(2021)\citenamefont {Boes},
  \citenamefont {Chang}, \citenamefont {Nguyen}, \citenamefont {Ren},
  \citenamefont {Bowers},\ and\ \citenamefont {Mitchell}}]{boes2021efficient}%
  \BibitemOpen
  \bibfield  {author} {\bibinfo {author} {\bibfnamefont {A.}~\bibnamefont
  {Boes}}, \bibinfo {author} {\bibfnamefont {L.}~\bibnamefont {Chang}},
  \bibinfo {author} {\bibfnamefont {T.}~\bibnamefont {Nguyen}}, \bibinfo
  {author} {\bibfnamefont {G.}~\bibnamefont {Ren}}, \bibinfo {author}
  {\bibfnamefont {J.}~\bibnamefont {Bowers}},\ and\ \bibinfo {author}
  {\bibfnamefont {A.}~\bibnamefont {Mitchell}},\ }\bibfield  {title} {\bibinfo
  {title} {Efficient second harmonic generation in lithium niobate on insulator
  waveguides and its pitfalls},\ }\href@noop {} {\bibfield  {journal} {\bibinfo
   {journal} {Journal of Physics: Photonics}\ }\textbf {\bibinfo {volume}
  {3}},\ \bibinfo {pages} {012008} (\bibinfo {year} {2021})}\BibitemShut
  {NoStop}%
\bibitem [{\citenamefont {Krasnokutska}\ \emph {et~al.}(2021)\citenamefont
  {Krasnokutska}, \citenamefont {Tambasco},\ and\ \citenamefont
  {Peruzzo}}]{krasnokutska2021submicron}%
  \BibitemOpen
  \bibfield  {author} {\bibinfo {author} {\bibfnamefont {I.}~\bibnamefont
  {Krasnokutska}}, \bibinfo {author} {\bibfnamefont {J.-L.~J.}\ \bibnamefont
  {Tambasco}},\ and\ \bibinfo {author} {\bibfnamefont {A.}~\bibnamefont
  {Peruzzo}},\ }\bibfield  {title} {\bibinfo {title} {Submicron domain
  engineering in periodically poled lithium niobate on insulator},\ }\href@noop
  {} {\bibfield  {journal} {\bibinfo  {journal} {arXiv preprint
  arXiv:2108.10839}\ } (\bibinfo {year} {2021})}\BibitemShut {NoStop}%
\bibitem [{\citenamefont {Conforti}\ \emph {et~al.}(2008)\citenamefont
  {Conforti}, \citenamefont {De~Angelis}, \citenamefont {Sapaev},\ and\
  \citenamefont {Assanto}}]{conforti2008pulse}%
  \BibitemOpen
  \bibfield  {author} {\bibinfo {author} {\bibfnamefont {M.}~\bibnamefont
  {Conforti}}, \bibinfo {author} {\bibfnamefont {C.}~\bibnamefont
  {De~Angelis}}, \bibinfo {author} {\bibfnamefont {U.~K.}\ \bibnamefont
  {Sapaev}},\ and\ \bibinfo {author} {\bibfnamefont {G.}~\bibnamefont
  {Assanto}},\ }\bibfield  {title} {\bibinfo {title} {Pulse shaping via
  backward second harmonic generation},\ }\href@noop {} {\bibfield  {journal}
  {\bibinfo  {journal} {Optics Express}\ }\textbf {\bibinfo {volume} {16}},\
  \bibinfo {pages} {2115} (\bibinfo {year} {2008})}\BibitemShut {NoStop}%
\bibitem [{\citenamefont {Iliew}\ \emph {et~al.}(2010)\citenamefont {Iliew},
  \citenamefont {Etrich}, \citenamefont {Pertsch}, \citenamefont {Lederer},\
  and\ \citenamefont {Kivshar}}]{Iliew2010Huge}%
  \BibitemOpen
  \bibfield  {author} {\bibinfo {author} {\bibfnamefont {R.}~\bibnamefont
  {Iliew}}, \bibinfo {author} {\bibfnamefont {C.}~\bibnamefont {Etrich}},
  \bibinfo {author} {\bibfnamefont {T.}~\bibnamefont {Pertsch}}, \bibinfo
  {author} {\bibfnamefont {F.}~\bibnamefont {Lederer}},\ and\ \bibinfo {author}
  {\bibfnamefont {Y.~S.}\ \bibnamefont {Kivshar}},\ }\bibfield  {title}
  {\bibinfo {title} {Huge enhancement of backward second-harmonic generation
  with slow light in photonic crystals},\ }\href
  {https://doi.org/10.1103/PhysRevA.81.023820} {\bibfield  {journal} {\bibinfo
  {journal} {Phys. Rev. A}\ }\textbf {\bibinfo {volume} {81}},\ \bibinfo
  {pages} {023820} (\bibinfo {year} {2010})}\BibitemShut {NoStop}%
\bibitem [{\citenamefont {Busacca}\ \emph {et~al.}(2014)\citenamefont
  {Busacca}, \citenamefont {Stivala}, \citenamefont {Curcio}, \citenamefont
  {Tomasino},\ and\ \citenamefont {Assanto}}]{busacca2014backward}%
  \BibitemOpen
  \bibfield  {author} {\bibinfo {author} {\bibfnamefont {A.~C.}\ \bibnamefont
  {Busacca}}, \bibinfo {author} {\bibfnamefont {S.}~\bibnamefont {Stivala}},
  \bibinfo {author} {\bibfnamefont {L.}~\bibnamefont {Curcio}}, \bibinfo
  {author} {\bibfnamefont {A.}~\bibnamefont {Tomasino}},\ and\ \bibinfo
  {author} {\bibfnamefont {G.}~\bibnamefont {Assanto}},\ }\bibfield  {title}
  {\bibinfo {title} {Backward frequency doubling of near infrared picosecond
  pulses},\ }\href@noop {} {\bibfield  {journal} {\bibinfo  {journal} {Optics
  express}\ }\textbf {\bibinfo {volume} {22}},\ \bibinfo {pages} {7544}
  (\bibinfo {year} {2014})}\BibitemShut {NoStop}%
\bibitem [{\citenamefont {Gu}\ \emph {et~al.}(1999)\citenamefont {Gu},
  \citenamefont {Makarov}, \citenamefont {Ding}, \citenamefont {Khurgin},\ and\
  \citenamefont {Risk}}]{Gu1999Backward}%
  \BibitemOpen
  \bibfield  {author} {\bibinfo {author} {\bibfnamefont {X.}~\bibnamefont
  {Gu}}, \bibinfo {author} {\bibfnamefont {M.}~\bibnamefont {Makarov}},
  \bibinfo {author} {\bibfnamefont {Y.~J.}\ \bibnamefont {Ding}}, \bibinfo
  {author} {\bibfnamefont {J.~B.}\ \bibnamefont {Khurgin}},\ and\ \bibinfo
  {author} {\bibfnamefont {W.~P.}\ \bibnamefont {Risk}},\ }\bibfield  {title}
  {\bibinfo {title} {Backward second-harmonic and third-harmonic generation in
  a periodically poled potassium titanyl phosphate waveguide},\ }\href
  {https://doi.org/10.1364/OL.24.000127} {\bibfield  {journal} {\bibinfo
  {journal} {Opt. Lett.}\ }\textbf {\bibinfo {volume} {24}},\ \bibinfo {pages}
  {127} (\bibinfo {year} {1999})}\BibitemShut {NoStop}%
\bibitem [{\citenamefont {Canalias}\ \emph {et~al.}(2005)\citenamefont
  {Canalias}, \citenamefont {Pasiskevicius}, \citenamefont {Fokine},\ and\
  \citenamefont {Laurell}}]{canalias2005backward}%
  \BibitemOpen
  \bibfield  {author} {\bibinfo {author} {\bibfnamefont {C.}~\bibnamefont
  {Canalias}}, \bibinfo {author} {\bibfnamefont {V.}~\bibnamefont
  {Pasiskevicius}}, \bibinfo {author} {\bibfnamefont {M.}~\bibnamefont
  {Fokine}},\ and\ \bibinfo {author} {\bibfnamefont {F.}~\bibnamefont
  {Laurell}},\ }\bibfield  {title} {\bibinfo {title} {Backward
  quasi-phase-matched second-harmonic generation in submicrometer periodically
  poled flux-grown k ti opo 4},\ }\href@noop {} {\bibfield  {journal} {\bibinfo
   {journal} {Applied Physics Letters}\ }\textbf {\bibinfo {volume} {86}},\
  \bibinfo {pages} {181105} (\bibinfo {year} {2005})}\BibitemShut {NoStop}%
\bibitem [{\citenamefont {Mutter}\ \emph {et~al.}(2022)\citenamefont {Mutter},
  \citenamefont {M{\o}lster}, \citenamefont {Zukauskas}, \citenamefont
  {Pasiskevicius},\ and\ \citenamefont {Canalias}}]{Mutter2022Third}%
  \BibitemOpen
  \bibfield  {author} {\bibinfo {author} {\bibfnamefont {P.}~\bibnamefont
  {Mutter}}, \bibinfo {author} {\bibfnamefont {K.~M.}\ \bibnamefont
  {M{\o}lster}}, \bibinfo {author} {\bibfnamefont {A.}~\bibnamefont
  {Zukauskas}}, \bibinfo {author} {\bibfnamefont {V.}~\bibnamefont
  {Pasiskevicius}},\ and\ \bibinfo {author} {\bibfnamefont {C.}~\bibnamefont
  {Canalias}},\ }\bibfield  {title} {\bibinfo {title} {Third order backward
  second-harmonic generation in periodically poled rb-doped ktp with a period
  of 317 nm},\ }in\ \href
  {http://opg.optica.org/abstract.cfm?URI=CLEO_SI-2022-SM2O.4} {\emph {\bibinfo
  {booktitle} {Conference on Lasers and Electro-Optics}}}\ (\bibinfo
  {publisher} {Optica Publishing Group},\ \bibinfo {year} {2022})\ p.\ \bibinfo
  {pages} {SM2O.4}\BibitemShut {NoStop}%
\bibitem [{\citenamefont {Stolt}\ \emph {et~al.}(2021)\citenamefont {Stolt},
  \citenamefont {Kim}, \citenamefont {H{\'e}ron}, \citenamefont {Vesala},
  \citenamefont {Yang}, \citenamefont {Mun}, \citenamefont {Kim}, \citenamefont
  {Huttunen}, \citenamefont {Czaplicki}, \citenamefont {Kauranen} \emph
  {et~al.}}]{stolt2021backward}%
  \BibitemOpen
  \bibfield  {author} {\bibinfo {author} {\bibfnamefont {T.}~\bibnamefont
  {Stolt}}, \bibinfo {author} {\bibfnamefont {J.}~\bibnamefont {Kim}}, \bibinfo
  {author} {\bibfnamefont {S.}~\bibnamefont {H{\'e}ron}}, \bibinfo {author}
  {\bibfnamefont {A.}~\bibnamefont {Vesala}}, \bibinfo {author} {\bibfnamefont
  {Y.}~\bibnamefont {Yang}}, \bibinfo {author} {\bibfnamefont {J.}~\bibnamefont
  {Mun}}, \bibinfo {author} {\bibfnamefont {M.}~\bibnamefont {Kim}}, \bibinfo
  {author} {\bibfnamefont {M.~J.}\ \bibnamefont {Huttunen}}, \bibinfo {author}
  {\bibfnamefont {R.}~\bibnamefont {Czaplicki}}, \bibinfo {author}
  {\bibfnamefont {M.}~\bibnamefont {Kauranen}}, \emph {et~al.},\ }\bibfield
  {title} {\bibinfo {title} {Backward phase-matched second-harmonic generation
  from stacked metasurfaces},\ }\href@noop {} {\bibfield  {journal} {\bibinfo
  {journal} {Physical Review Letters}\ }\textbf {\bibinfo {volume} {126}},\
  \bibinfo {pages} {033901} (\bibinfo {year} {2021})}\BibitemShut {NoStop}%
\bibitem [{\citenamefont {Lan}\ \emph {et~al.}(2015)\citenamefont {Lan},
  \citenamefont {Kang}, \citenamefont {Schoen}, \citenamefont {Rodrigues},
  \citenamefont {Cui}, \citenamefont {Brongersma},\ and\ \citenamefont
  {Cai}}]{lan2015backward}%
  \BibitemOpen
  \bibfield  {author} {\bibinfo {author} {\bibfnamefont {S.}~\bibnamefont
  {Lan}}, \bibinfo {author} {\bibfnamefont {L.}~\bibnamefont {Kang}}, \bibinfo
  {author} {\bibfnamefont {D.~T.}\ \bibnamefont {Schoen}}, \bibinfo {author}
  {\bibfnamefont {S.~P.}\ \bibnamefont {Rodrigues}}, \bibinfo {author}
  {\bibfnamefont {Y.}~\bibnamefont {Cui}}, \bibinfo {author} {\bibfnamefont
  {M.~L.}\ \bibnamefont {Brongersma}},\ and\ \bibinfo {author} {\bibfnamefont
  {W.}~\bibnamefont {Cai}},\ }\bibfield  {title} {\bibinfo {title} {Backward
  phase-matching for nonlinear optical generation in negative-index
  materials},\ }\href@noop {} {\bibfield  {journal} {\bibinfo  {journal}
  {Nature Materials}\ }\textbf {\bibinfo {volume} {14}},\ \bibinfo {pages}
  {807} (\bibinfo {year} {2015})}\BibitemShut {NoStop}%
\bibitem [{\citenamefont {Liu}\ \emph {et~al.}(2018)\citenamefont {Liu},
  \citenamefont {Wu}, \citenamefont {Zhang}, \citenamefont {Li}, \citenamefont
  {Zhang},\ and\ \citenamefont {Luo}}]{liu2018backward}%
  \BibitemOpen
  \bibfield  {author} {\bibinfo {author} {\bibfnamefont {L.}~\bibnamefont
  {Liu}}, \bibinfo {author} {\bibfnamefont {L.}~\bibnamefont {Wu}}, \bibinfo
  {author} {\bibfnamefont {J.}~\bibnamefont {Zhang}}, \bibinfo {author}
  {\bibfnamefont {Z.}~\bibnamefont {Li}}, \bibinfo {author} {\bibfnamefont
  {B.}~\bibnamefont {Zhang}},\ and\ \bibinfo {author} {\bibfnamefont
  {Y.}~\bibnamefont {Luo}},\ }\bibfield  {title} {\bibinfo {title} {Backward
  phase matching for second harmonic generation in negative-index conformal
  surface plasmonic metamaterials},\ }\href@noop {} {\bibfield  {journal}
  {\bibinfo  {journal} {Advanced Science}\ }\textbf {\bibinfo {volume} {5}},\
  \bibinfo {pages} {1800661} (\bibinfo {year} {2018})}\BibitemShut {NoStop}%
\bibitem [{\citenamefont {Krol}\ \emph {et~al.}(1991)\citenamefont {Krol},
  \citenamefont {Broer}, \citenamefont {Nelson}, \citenamefont {Stolen},
  \citenamefont {Tom},\ and\ \citenamefont {Pleibel}}]{krol1991seeded}%
  \BibitemOpen
  \bibfield  {author} {\bibinfo {author} {\bibfnamefont {D.~M.}\ \bibnamefont
  {Krol}}, \bibinfo {author} {\bibfnamefont {M.~M.}\ \bibnamefont {Broer}},
  \bibinfo {author} {\bibfnamefont {K.~T.}\ \bibnamefont {Nelson}}, \bibinfo
  {author} {\bibfnamefont {R.~H.}\ \bibnamefont {Stolen}}, \bibinfo {author}
  {\bibfnamefont {H.~W.~K.}\ \bibnamefont {Tom}},\ and\ \bibinfo {author}
  {\bibfnamefont {W.}~\bibnamefont {Pleibel}},\ }\bibfield  {title} {\bibinfo
  {title} {Seeded second-harmonic generation in optical fibers: the effect of
  phase fluctuations},\ }\href {https://doi.org/10.1364/OL.16.000211}
  {\bibfield  {journal} {\bibinfo  {journal} {Opt. Lett.}\ }\textbf {\bibinfo
  {volume} {16}},\ \bibinfo {pages} {211} (\bibinfo {year} {1991})}\BibitemShut
  {NoStop}%
\bibitem [{\citenamefont {Yakar}\ \emph {et~al.}(2022)\citenamefont {Yakar},
  \citenamefont {Nitiss}, \citenamefont {Hu},\ and\ \citenamefont
  {Brès}}]{yakar2022coherent}%
  \BibitemOpen
  \bibfield  {author} {\bibinfo {author} {\bibfnamefont {O.}~\bibnamefont
  {Yakar}}, \bibinfo {author} {\bibfnamefont {E.}~\bibnamefont {Nitiss}},
  \bibinfo {author} {\bibfnamefont {J.}~\bibnamefont {Hu}},\ and\ \bibinfo
  {author} {\bibfnamefont {C.-S.}\ \bibnamefont {Brès}},\ }\bibfield  {title}
  {\bibinfo {title} {Generalized coherent photogalvanic effect in coherently
  seeded waveguides},\ }\href
  {https://doi.org/https://doi.org/10.1002/lpor.202200294} {\bibfield
  {journal} {\bibinfo  {journal} {Laser \& Photonics Reviews}\ }\textbf
  {\bibinfo {volume} {n/a}},\ \bibinfo {pages} {2200294} (\bibinfo {year}
  {2022})}\BibitemShut {NoStop}%
\bibitem [{\citenamefont {Zabelich}\ \emph
  {et~al.}(2022{\natexlab{a}})\citenamefont {Zabelich}, \citenamefont {Nitiss},
  \citenamefont {Stroganov},\ and\ \citenamefont
  {Br{\`e}s}}]{zabelich2022linear}%
  \BibitemOpen
  \bibfield  {author} {\bibinfo {author} {\bibfnamefont {B.}~\bibnamefont
  {Zabelich}}, \bibinfo {author} {\bibfnamefont {E.}~\bibnamefont {Nitiss}},
  \bibinfo {author} {\bibfnamefont {A.}~\bibnamefont {Stroganov}},\ and\
  \bibinfo {author} {\bibfnamefont {C.-S.}\ \bibnamefont {Br{\`e}s}},\
  }\bibfield  {title} {\bibinfo {title} {Linear electro-optic effect in silicon
  nitride waveguides enabled by electric-field poling},\ }\href@noop {}
  {\bibfield  {journal} {\bibinfo  {journal} {ACS photonics}\ } (\bibinfo
  {year} {2022}{\natexlab{a}})}\BibitemShut {NoStop}%
\bibitem [{\citenamefont {Zabelich}\ \emph
  {et~al.}(2022{\natexlab{b}})\citenamefont {Zabelich}, \citenamefont {Nitiss},
  \citenamefont {Stroganov},\ and\ \citenamefont
  {Br{\`e}s}}]{zabelich2022electric}%
  \BibitemOpen
  \bibfield  {author} {\bibinfo {author} {\bibfnamefont {B.}~\bibnamefont
  {Zabelich}}, \bibinfo {author} {\bibfnamefont {E.}~\bibnamefont {Nitiss}},
  \bibinfo {author} {\bibfnamefont {A.}~\bibnamefont {Stroganov}},\ and\
  \bibinfo {author} {\bibfnamefont {C.-S.}\ \bibnamefont {Br{\`e}s}},\
  }\bibfield  {title} {\bibinfo {title} {Electric-field poling of silicon
  nitride waveguides for the linear phase modulation},\ }in\ \href@noop {}
  {\emph {\bibinfo {booktitle} {Nonlinear Optics and its Applications 2022}}},\
  Vol.\ \bibinfo {volume} {12143}\ (\bibinfo {organization} {SPIE},\ \bibinfo
  {year} {2022})\ pp.\ \bibinfo {pages} {48--54}\BibitemShut {NoStop}%
\bibitem [{\citenamefont {Fejer}\ \emph {et~al.}(1992)\citenamefont {Fejer},
  \citenamefont {Magel}, \citenamefont {Jundt},\ and\ \citenamefont
  {Byer}}]{fejer1992quasi}%
  \BibitemOpen
  \bibfield  {author} {\bibinfo {author} {\bibfnamefont {M.~M.}\ \bibnamefont
  {Fejer}}, \bibinfo {author} {\bibfnamefont {G.}~\bibnamefont {Magel}},
  \bibinfo {author} {\bibfnamefont {D.~H.}\ \bibnamefont {Jundt}},\ and\
  \bibinfo {author} {\bibfnamefont {R.~L.}\ \bibnamefont {Byer}},\ }\bibfield
  {title} {\bibinfo {title} {Quasi-phase-matched second harmonic generation:
  tuning and tolerances},\ }\href@noop {} {\bibfield  {journal} {\bibinfo
  {journal} {IEEE Journal of Quantum Electronics}\ }\textbf {\bibinfo {volume}
  {28}},\ \bibinfo {pages} {2631} (\bibinfo {year} {1992})}\BibitemShut
  {NoStop}%
\bibitem [{\citenamefont {Nitiss}\ \emph
  {et~al.}(2020{\natexlab{b}})\citenamefont {Nitiss}, \citenamefont {Yakar},
  \citenamefont {Stroganov},\ and\ \citenamefont
  {Br{\`e}s}}]{nitiss2020highly}%
  \BibitemOpen
  \bibfield  {author} {\bibinfo {author} {\bibfnamefont {E.}~\bibnamefont
  {Nitiss}}, \bibinfo {author} {\bibfnamefont {O.}~\bibnamefont {Yakar}},
  \bibinfo {author} {\bibfnamefont {A.}~\bibnamefont {Stroganov}},\ and\
  \bibinfo {author} {\bibfnamefont {C.-S.}\ \bibnamefont {Br{\`e}s}},\
  }\bibfield  {title} {\bibinfo {title} {Highly tunable second-harmonic
  generation in all-optically poled silicon nitride waveguides},\ }\href@noop
  {} {\bibfield  {journal} {\bibinfo  {journal} {Optics Letters}\ }\textbf
  {\bibinfo {volume} {45}},\ \bibinfo {pages} {1958} (\bibinfo {year}
  {2020}{\natexlab{b}})}\BibitemShut {NoStop}%
\end{thebibliography}%


\begin{thebibliography}{3}%
\makeatletter
\providecommand \@ifxundefined [1]{%
 \@ifx{#1\undefined}
}%
\providecommand \@ifnum [1]{%
 \ifnum #1\expandafter \@firstoftwo
 \else \expandafter \@secondoftwo
 \fi
}%
\providecommand \@ifx [1]{%
 \ifx #1\expandafter \@firstoftwo
 \else \expandafter \@secondoftwo
 \fi
}%
\providecommand \natexlab [1]{#1}%
\providecommand \enquote  [1]{``#1''}%
\providecommand \bibnamefont  [1]{#1}%
\providecommand \bibfnamefont [1]{#1}%
\providecommand \citenamefont [1]{#1}%
\providecommand \href@noop [0]{\@secondoftwo}%
\providecommand \href [0]{\begingroup \@sanitize@url \@href}%
\providecommand \@href[1]{\@@startlink{#1}\@@href}%
\providecommand \@@href[1]{\endgroup#1\@@endlink}%
\providecommand \@sanitize@url [0]{\catcode `\\12\catcode `\$12\catcode
  `\&12\catcode `\#12\catcode `\^12\catcode `\_12\catcode `\%12\relax}%
\providecommand \@@startlink[1]{}%
\providecommand \@@endlink[0]{}%
\providecommand \url  [0]{\begingroup\@sanitize@url \@url }%
\providecommand \@url [1]{\endgroup\@href {#1}{\urlprefix }}%
\providecommand \urlprefix  [0]{URL }%
\providecommand \Eprint [0]{\href }%
\providecommand \doibase [0]{https://doi.org/}%
\providecommand \selectlanguage [0]{\@gobble}%
\providecommand \bibinfo  [0]{\@secondoftwo}%
\providecommand \bibfield  [0]{\@secondoftwo}%
\providecommand \translation [1]{[#1]}%
\providecommand \BibitemOpen [0]{}%
\providecommand \bibitemStop [0]{}%
\providecommand \bibitemNoStop [0]{.\EOS\space}%
\providecommand \EOS [0]{\spacefactor3000\relax}%
\providecommand \BibitemShut  [1]{\csname bibitem#1\endcsname}%
\let\auto@bib@innerbib\@empty
\bibitem [{\citenamefont {Zabelich}\ \emph {et~al.}(2022)\citenamefont
  {Zabelich}, \citenamefont {Nitiss}, \citenamefont {Stroganov},\ and\
  \citenamefont {Br{\`e}s}}]{zabelich2022linear}%
  \BibitemOpen
  \bibfield  {author} {\bibinfo {author} {\bibfnamefont {B.}~\bibnamefont
  {Zabelich}}, \bibinfo {author} {\bibfnamefont {E.}~\bibnamefont {Nitiss}},
  \bibinfo {author} {\bibfnamefont {A.}~\bibnamefont {Stroganov}},\ and\
  \bibinfo {author} {\bibfnamefont {C.-S.}\ \bibnamefont {Br{\`e}s}},\
  }\bibfield  {title} {\bibinfo {title} {Linear electro-optic effect in silicon
  nitride waveguides enabled by electric-field poling},\ }\href@noop {}
  {\bibfield  {journal} {\bibinfo  {journal} {ACS photonics}\ } (\bibinfo
  {year} {2022})}\BibitemShut {NoStop}%
\bibitem [{\citenamefont {Yakar}\ \emph {et~al.}(2022)\citenamefont {Yakar},
  \citenamefont {Nitiss}, \citenamefont {Hu},\ and\ \citenamefont
  {Brès}}]{yakar2022coherent}%
  \BibitemOpen
  \bibfield  {author} {\bibinfo {author} {\bibfnamefont {O.}~\bibnamefont
  {Yakar}}, \bibinfo {author} {\bibfnamefont {E.}~\bibnamefont {Nitiss}},
  \bibinfo {author} {\bibfnamefont {J.}~\bibnamefont {Hu}},\ and\ \bibinfo
  {author} {\bibfnamefont {C.-S.}\ \bibnamefont {Brès}},\ }\bibfield  {title}
  {\bibinfo {title} {Generalized coherent photogalvanic effect in coherently
  seeded waveguides},\ }\href
  {https://doi.org/https://doi.org/10.1002/lpor.202200294} {\bibfield
  {journal} {\bibinfo  {journal} {Laser \& Photonics Reviews}\ }\textbf
  {\bibinfo {volume} {n/a}},\ \bibinfo {pages} {2200294} (\bibinfo {year}
  {2022})}\BibitemShut {NoStop}%
\bibitem [{coh(1998)}]{cohen1998atom}%
  \BibitemOpen
  \bibinfo {title} {Appendix},\ in\ \href
  {https://doi.org/https://doi.org/10.1002/9783527617197.app1} {\emph {\bibinfo
  {booktitle} {Atom—Photon Interactions}}}\ (\bibinfo  {publisher} {John
  Wiley \& Sons, Ltd},\ \bibinfo {year} {1998})\ pp.\ \bibinfo {pages}
  {621--639},\ \Eprint
  {https://arxiv.org/abs/https://onlinelibrary.wiley.com/doi/pdf/10.1002/9783527617197.app1}
  {https://onlinelibrary.wiley.com/doi/pdf/10.1002/9783527617197.app1}
  \BibitemShut {NoStop}%
\end{thebibliography}%

\end{document}


\title{Supplementary Information for: Integrated Backward Second-Harmonic Generation Through Optically Induced Quasi-Phase Matching}

\author{Ozan Yakar$^1$}
\author{Edgars Nitiss$^1$}%
\author{Jianqi Hu$^1$}%
\author{Camille-Sophie Br\`{e}s$^{1,}$}%
 \email{camille.bres@epfl.ch}
\affiliation{$^1$\'Ecole Polytechnique Fédérale de Lausanne (EPFL), Photonic Systems Laboratory (PHOSL), Lausanne CH-1015, Switzerland }

\date{\today}

\maketitle


\renewcommand{\theequation}{S\arabic{equation}}
\renewcommand{\thefigure}{S\arabic{figure}}
\renewcommand{\thetable}{S\arabic{table}}

\tableofcontents
\section{Derivation of Static Charge Equations}
In this supplementary note, we derive the electric field in a waveguide with periodic charge distribution along the propagation direction. The electric field ($\vec{E}$) at position $\vec{r}_0$ can be found from Gauss' Law as 
\begin{equation}
    \vec{E}(\vec{r}_0)=\frac{1}{4\pi\epsilon}\iiint \limits_{V} d^3\vec{r} \frac{\rho(\vec{r})}{|\vec{r}-\vec{r}_0|^3}(\vec{r}-\vec{r}_0)
\end{equation}
Then $E_y(x_0,y_0,z_0)$ can be found as
\begin{equation}
    E_y (x_0, y_0, z_0)=\frac{1}{4 \pi \epsilon} \int \limits_{-\infty}^{+\infty} d x\int \limits_{-\infty}^{+\infty} d y\int \limits_{-\infty}^{+\infty} d z  \frac{\rho\left(x+x_0, y+y_0,z+z_0\right) y}{\left(x^2+y+z^2\right)^{3 / 2}}
\end{equation}
$\rho(x,y,z)=\rho_0(x,y)\cos(\Delta k z)$ and just the integral with respect to z for $\Delta k\neq 0$ can be represented as,
\begin{equation}
    \Delta k^2\int \limits_{-\infty}^{+\infty} \Delta k d z  \frac{\cos(\Delta k (z+z_0)) }{\Delta k^3\left(x^2+y+z^2\right)^{3 / 2}}=2\Delta k  \cos \left(\Delta k z_0\right) \frac{K_1\left(\Delta k \sqrt{x^2+y^2}\right)}{\sqrt{x^2+y^2}}    
\end{equation}

If $\Delta k\neq0$,
\begin{equation}
    E_y\left(x_0, y_0, z_0, \Delta k\right)=\frac{\Delta k \cos \left(\Delta k z_0\right)}{2 \pi \epsilon} \int \limits_{-\infty}^{+\infty} \int \limits_{-\infty}^{+\infty} d x d y   \frac{K_1\left(\Delta k \sqrt{x^2+y^2}\right)}{\sqrt{x^2+y^2}}y \rho_0(x+x_0,y+y_0)
    \label{eq:Eydk}
\end{equation}
If $\Delta k=0$, 

\begin{equation}
    E_y\left(x_0, y_0, z_0\right)=\frac{1}{2 \pi \epsilon} \int \limits_{-\infty}^{+\infty}\int \limits_{-\infty}^{+\infty} d x d y \frac{y \rho_0\left(x+x_0, y+y_0\right)} {\left(x^2+y^2\right)^2}
\end{equation}

The eq. \eqref{eq:Eydk} rewrites as
\begin{equation}
    E_y\left(x_0, y_0, z_0, \Delta k\right)=\frac{\Delta k \cos \left(\Delta k z_0\right)}{2 \pi \epsilon} \int \limits_{-\infty}^{+\infty} \int \limits_{-\infty}^{+\infty} d x d y (y-y_0)  \frac{K_1\left(\Delta k \sqrt{(x-x_0)^2+(y-y_0)^2}\right)}{\sqrt{(x-x_0)^2+(y-y_0)^2}} \rho_0(x,y)
\end{equation}
This can be rewritten as
\begin{equation}
    E_y\left(x_0, y_0, \Delta k\right)=-\frac{\Delta k \cos \left(\Delta k z_0\right)}{2 \pi \epsilon} \int \limits_{-\infty}^{+\infty} \int \limits_{-\infty}^{+\infty} d x d y (y_0-y)  \frac{K_1\left(\Delta k \sqrt{(x_0-x)^2+(y_0-y)^2}\right)}{\sqrt{(x_0-x)^2+(y_0-y)^2}} \rho_0(x,y)
\end{equation}
This is a convolution expression. 
\begin{equation}
    E_y\left(x_0, y_0, \Delta k\right)=-\frac{\Delta k \cos \left(\Delta k z_0\right)}{2 \pi \epsilon} \left[ y  \frac{K_1\left(\Delta k \sqrt{x^2+y^2}\right)}{\sqrt{x^2+y^2}} * \rho_0(x,y) \right]
\end{equation} 

\section{The Change of Time Constant with Grating Period}

In this supplementary note, we discuss the effect of diffusion on the time constants. The conservation of charge writes 
\begin{equation}
    \partial_t \rho = -\vec{\nabla} \cdot \vec{j}_{ph} - \frac{\sigma}{\epsilon}\rho+ D \nabla^2 \rho
\end{equation}
where $\vec{j}_{ph}$ photocurrent, $\sigma$ is the photoconductivity and $D$ is the diffusivity. The last term is due to diffusion current and, for the charge distribution above, it becomes $-D\Delta k^2\rho + D\nabla_T^2 \rho$, where $\nabla_T^2$ is the transverse Laplacian operator. For BSHG, change in longitudinal direction of charge density is much more than the transverse direction. Hence, the conservation of charge can be approximated as $\partial_t \rho \approx -\vec{\nabla} \cdot \vec{j}_{ph} - \frac{1}{\tau} \rho$. Then, the time constant becomes 
\begin{equation}
    \tau \approx \frac{1}{\frac{\sigma}{\epsilon} + D \Delta k^2}
    \label{eq:tauall}
\end{equation}
Using the diffusivity value from Ref. \cite{zabelich2022linear}, the latter term in the denominator of \eqref{eq:tauall} becomes comparable to the values found from Ref. \cite{yakar2022coherent} for very short poling periods reducing the time constant. Using the values from \cite{zabelich2022linear}, diffusion becomes important when the poling times are higher than 1000 seconds. So effect of diffusion is not significant for the powers we work with.
\section{Effect of Grating Period on the Growth Rate}

In this supplementary note, we derive the dynamics of second-harmonic generation for an arbitrary grating period and we find the initial growth rate of generated second-harmonic. 

Conservation of charge can be re-written as: 
\begin{equation}
    \frac{\partial\rho}{\partial t}= -\nabla \cdot \vec{j} - \frac{\rho}{\tau}
    \label{eq:conccharge}
\end{equation}
$\rho$ relates to electric field with $-\frac{i \hat{k}}{\epsilon k}$ in the Fourier domain \cite{cohen1998atom}. We assume the current is in the y-direction. Hence taking the spatial Fourier transform of above equation and multiplying with the previous expression, we find the y component of the electric field as
\begin{equation}
    \frac{\partial\hat{E}_y}{\partial t}= -\frac{k_y^2 \hat{{j}}}{ \epsilon (k_x^2+k_y^2+k_z^2)} - \frac{\hat{E}_y}{\tau}
\end{equation}
where $\hat{k}=\vec{k}/k$ and the hat on the other variables denote the spatial Fourier transform. The current is chosen to be in the y-direction. Using $\chi^{(2)}=3\chi^{(3)}E_y$ and $\chi^{(3)}(z)=\chi^{(3)}u(z)$ where $u(z)$ is the unit-step function, we have
\begin{equation}
    \frac{\partial\hat{{\chi}}^{(2)}}{\partial t}= 3\chi^{(3)}\frac{\mathcal{F}(u(z))}{\sqrt{2\pi}}\ast\frac{k_y^2 \hat{{j}}}{ \epsilon (k_x^2+k_y^2+k_z^2)} - \frac{\hat{{\chi}}^{(2)}}{\tau}
\end{equation}
Here, the convolution theorem is $\mathcal{F}[f(z)\ast g(z)]=\sqrt{2\pi}f(k)g(k)$. Taking the inverse Fourier transform along the propagation direction and using $j_{ph}=\beta (E_{\omega}^*)^2E_{2\omega}$ and $E_{q\omega}\approx U_{q\omega}(x,y)A_{q\omega}(z,t)$ where q is 1 or 2 with the normalization $\iint dxdy\, |U_{q \omega}(x,y)|^{2}=1$ and using undepleted pump approximation similar to Ref. \cite{yakar2022coherent}, we have
\begin{equation}
    \frac{\partial\hat{{\chi}}^{(2)}(k_x,k_y,z)}{\partial t}= 3\chi^{(3)}\beta e^{-i\psi} (A_{\omega}^*)^2u(z)\left[\frac{ k_y^2 e^{-\sqrt{k_x^2+k_y^2}|z|}}{2 \epsilon \sqrt{k_x^2+k_y^2}}\hat{j}_{\rm mode}\ast A_{2\omega}e^{i\Delta k z}\right]  - \frac{\hat{{\chi}}^{(2)}}{\tau}
\end{equation}
where $\hat{j}_{\rm mode}=\mathcal{F}[(U_{\omega}^*)^2U_{2\omega}]$, and $\ast$ denotes convolution. We can write $\bar{\chi}^{(2)}=\chi^{(2)}e^{t/\tau}$ and $\bar{A}=A_{2\omega}e^{t/\tau}$. So, we have
\begin{equation}
    \frac{\partial\hat{\bar{\chi}}^{(2)}(k_x,k_y,z)}{\partial t}= 3\chi^{(3)}\beta e^{-i\psi}(A_{\omega}^*)^2u(z)\left[\frac{ k_y^2 e^{-\sqrt{k_x^2+k_y^2}|z|}}{2 \epsilon \sqrt{k_x^2+k_y^2}}\hat{j}_{\rm mode}\ast \bar{A}e^{i\Delta k z}\right]
    \label{eq:chi2ft}
\end{equation}

The partial derivative of $A_{2\omega}$ with respect to t using the second-harmonic generation equation with slowly varying envelope approximation writes as
\begin{equation}
    \frac{\partial^2 \bar{A}}{\partial z \partial t}= \frac{i\omega}{2n_{2\omega}c} A_{\omega}^2 e^{-i\Delta k z} \iint dx dy U_{2\omega}^*\frac{\partial{\bar{\chi}}^{(2)}}{\partial t} U_{\omega}^2
    \label{eq:shg}
\end{equation}
plugging eq. \eqref{eq:chi2ft} into eq. \eqref{eq:shg} and using the Parseval identity we have
\begin{equation}
    \frac{\partial^2 \bar{A}}{\partial z \partial t}= \frac{i 3\chi^{(3)}\omega}{2\epsilon n_{2\omega}c} \beta |A_{\omega}|^4 e^{-i\Delta k z}e^{-i\psi}u(z) \iint dk_x dk_y \frac{k_y^2 e^{-\sqrt{k_x^2+k_y^2}|z|}}{2 \sqrt{k_x^2+k_y^2}} |\hat{j}_{\rm mode}|^2\ast \bar{A}e^{i\Delta k z}
\end{equation}

For BSHG, as the generates SH is few orders of magnitude smaller than the seed we can assume the total SH is constant. Hence we replace $\bar{A}$ on the right hand side of above equation with $\bar{A}^{\rm s}$ and left hand side with $\bar{A}^{\rm g}$. Defining $M=\frac{i3\omega\chi^{(3)}}{2n_{2\omega}c\epsilon}\beta|A_{\omega}|^4e^{-i\psi}$, we have
\begin{figure}[h!]
    \centering
    \includegraphics[width=1\linewidth]{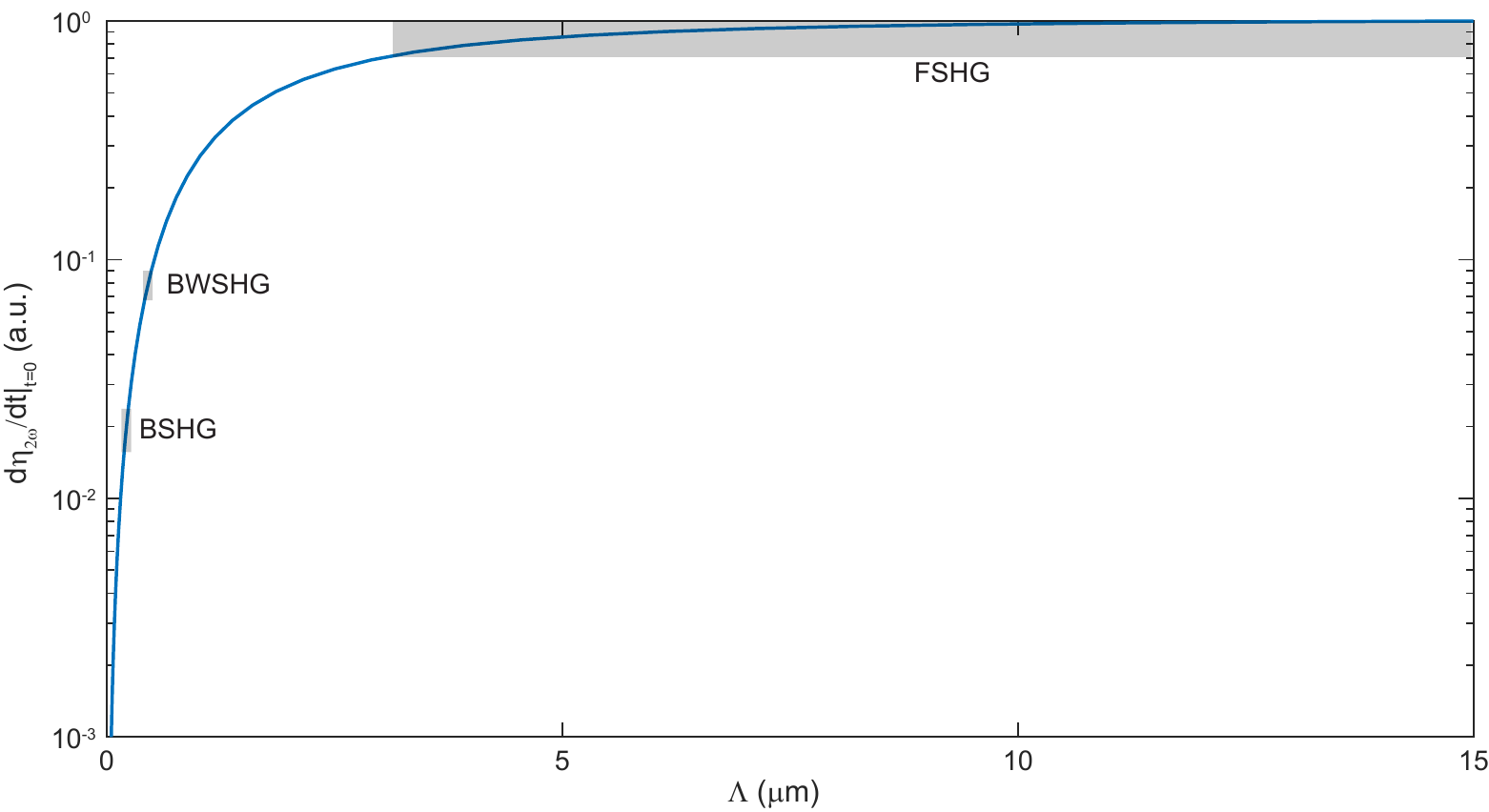}
    \caption{Simulated initial growth rate for different grating periods corresponding to BSHG, Backward-wave SHG (BWSHG) and FSHG.}
    \label{fig:gammaplot}
\end{figure}
\begin{equation}
    \frac{\partial^2 \bar{A}^{\rm g}}{\partial z \partial t}= M e^{-i\Delta k z}u(z) \iint dk_x dk_y \frac{k_y^2 e^{-\sqrt{k_x^2+k_y^2}|z|}}{2 \sqrt{k_x^2+k_y^2}} |\hat{j}_{\rm mode}|^2\ast \bar{A}^{\rm s}e^{i\Delta k z}
\end{equation}
The convolution expression can be evaluated and one finds
\begin{equation}
    \frac{\partial^2 \bar{A}^{\rm g}}{\partial z \partial t}= M\bar{A}^{\rm s} u(z) \iint dk_x dk_y \frac{k_y^2}{2 \sqrt{k_x^2+k_y^2}} |\hat{j}_{\rm mode}|^2\left(\frac{2\sqrt{k_x^2+k_y^2}}{k_x^2+k_y^2+\Delta k^2}-\frac{e^{-\sqrt{k_x^2+k_y^2}z}e^{-i\Delta k z}}{\sqrt{k_x^2+k_y^2}+i\Delta k} \right)
    \label{eq:gov}
\end{equation}
Hence, the generated SH for BSHG becomes
\begin{equation}
    A_{2\omega}^{\rm g}=A^{\rm s}_{2\omega}(1-e^{-t/\tau})M\tau \iint dk_x dk_y \frac{k_y^2}{\sqrt{k_x^2+k_y^2}}|\hat{j}_{\rm mode}|^2 \left(\frac{\sqrt{k_x^2+k_y^2}z}{k_x^2+k_y^2+\Delta k^2}+\frac{1-e^{-\sqrt{k_x^2+k_y^2}z}e^{-i\Delta k z}}{2(\sqrt{k_x^2+k_y^2}+i\Delta k)^2} \right)
\end{equation}
Above integral holds the information on waveguide dimensions and grating period. In the initial condition the assumptions made holds for any grating period and in order to compare FSHG and BSHG initial growth rate can be calculated. The initial growth rate becomes
\begin{equation}
    \left.\frac{\partial A_{2\omega}^{\rm g}}{\partial t}\right|_{t=0}=A^{\rm s}_{2\omega}M \iint dk_x dk_y \frac{k_y^2}{\sqrt{k_x^2+k_y^2}}|\hat{j}_{\rm mode}|^2 \left(\frac{\sqrt{k_x^2+k_y^2}z}{k_x^2+k_y^2+\Delta k^2}+\frac{1-e^{-\sqrt{k_x^2+k_y^2}z}e^{-i\Delta k z}}{2(\sqrt{k_x^2+k_y^2}+i\Delta k)^2} \right)
    \label{eq:BSHGtime}
\end{equation}

From Leibniz's rule, the initial growth rate as well as the efficiency decreases with increasing wavevector mismatch. The integral reduces to overlap integral times length for long waveguide lengths much longer than the waveguide crossections. In Fig. \ref{fig:gammaplot}, the initial growth rate is shown. 
\begin{figure}[h!]
    \centering
    \includegraphics[width=1\linewidth]{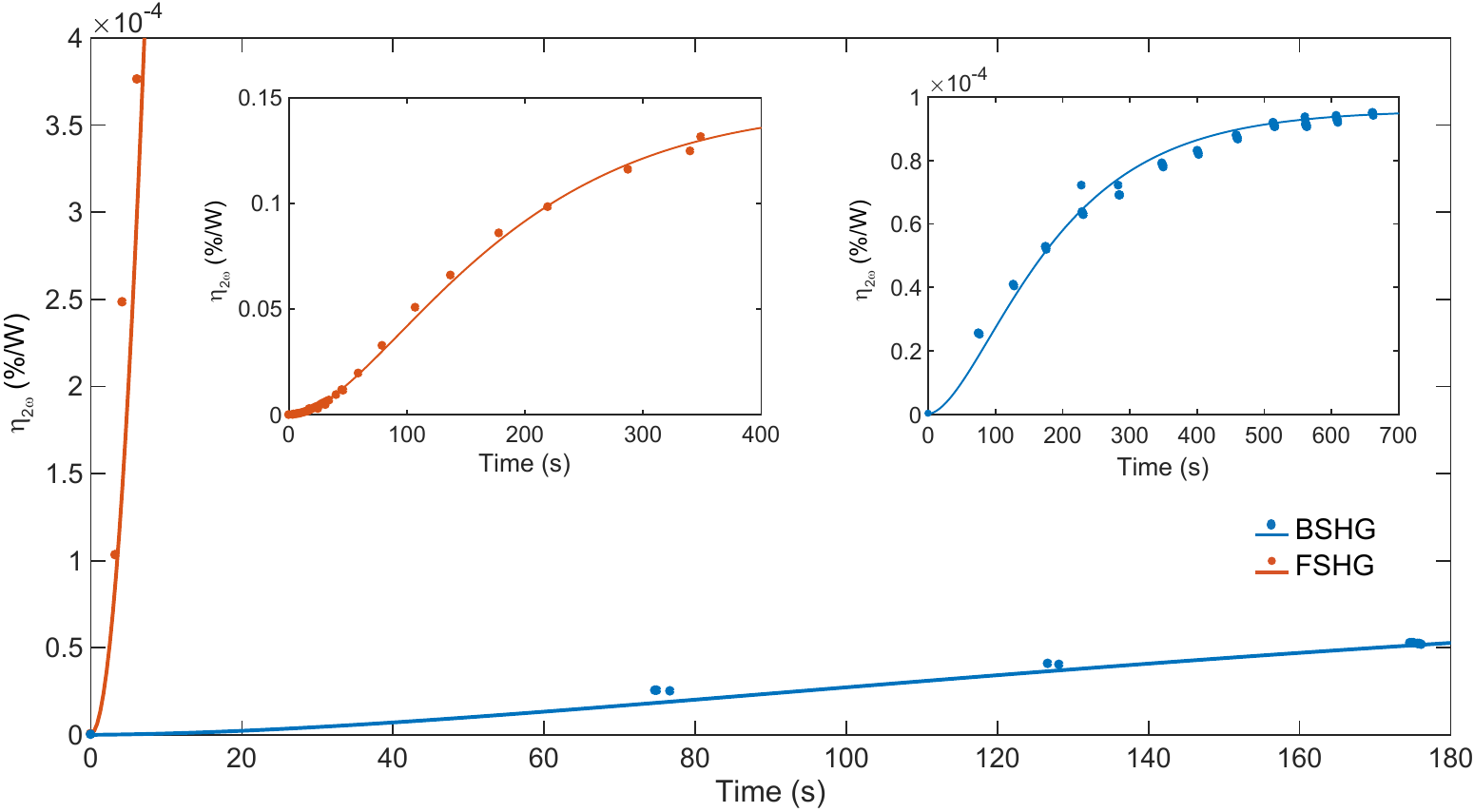}
    \caption{Experimental SHG CEs during AOP (dotted) for BSHG (blue) and FSHG (orange) and fits (solid) using
Equation from ref. \cite{yakar2022coherent} and \eqref{eq:BSHGtime} for constant peak SH seed power of $117\pm17$ mW
and and peak pump power of $8.4\pm0.1$ W. Insets: Entire AOP trace of BSHG (right) and FSHG (left).}
    \label{fig:comparebf}
\end{figure}

\section{Comparison of Experimental Growth Rates of FSHG and BSHG}

There has been interest on the how the material constants affect the AOP dynamics. In this section, we demonstrate the experimental growth rates of FSHG and BSHG enabled by AOP and confirm the effect of diffusivity on time constants is not significant for BSHG, as well. The AOP time trace for FSHG and BSHG are shown in Fig. \ref{fig:comparebf} (inset, left and right) respectively. The initial growth rates are shown in Fig. \ref{fig:comparebf} and the extracted initial growth rates vary by approximately $100$ times. This well agrees with the theoretical expectations shown in Fig. 2(b). The time constants are extracted to be the same ($135$ s) and much shorter than the time constant expected by diffusivity which is in the order of $1000$ s. For our experimental conditions, we, therefore, observe that photoconductivity dominates as the limiting factor.

\bibliography{apssamp}